\documentclass[]{siamart1116}

\floatstyle{plaintop}
\restylefloat{table}

\usepackage{amsfonts}
\usepackage{amsmath}
\usepackage{graphicx}
\usepackage{epstopdf}
\usepackage{algorithm,algpseudocode}
\usepackage{mathtools}
\usepackage{multirow}
\usepackage{booktabs}
\usepackage{bm}
\usepackage[caption=false]{subfig}
\usepackage{float}

\ifpdf
  \DeclareGraphicsExtensions{.eps,.pdf,.png,.jpg}
\else
  \DeclareGraphicsExtensions{.eps}
\fi

\numberwithin{theorem}{section}

\newcommand{\TheTitle}{Nonlocal Machine Learning of Micro-Structural Defect Evolutions in Crystalline Materials} 
\newcommand{\TheAuthors}{E. A. Barros de Moraes, M. D'Elia and M. Zayernouri}

\headers{}{}

\author{
  Eduardo A. Barros de Moraes\thanks{Department of Mechanical Engineering \& Department of Computational Mathematics, Science and Engineering, Michigan State University, 428 S Shaw Ln, East Lansing, MI 48824, USA.}
  \and
  Marta D'Elia\thanks{Computational Science and Analysis, Sandia National Laboratories, CA, USA.}
  \and
  Mohsen Zayernouri \thanks{Department of Mechanical Engineering \& Department of Statistics and Probability, Michigan State University, 428 S Shaw Ln, East Lansing, MI 48824, USA, Corresponding Author; \email{zayern@msu.edu}}
}

\title{{\TheTitle}\thanks{EABDM and MZ were supported by the US ARO Young Investigator Program Award (W911NF-19-1-0444) and partially by the National Science Foundation Award (DMS-1923201). The HPC resources and services were provided by the Institute for Cyber-Enabled Research (ICER) at Michigan State University. MD is supported by the U.S. Department of Energy, Office of Advanced Scientific Computing Research under the Collaboratory on Mathematics and Physics-Informed Learning Machines for Multiscale and Multiphysics Problems (PhILMs) project. Sandia National Laboratories is a multi-mission laboratory managed and operated by National Technology and Engineering Solutions of Sandia, LLC., a wholly owned subsidiary of Honeywell International, Inc., for the U.S. Department of Energy’s National Nuclear Security Administration under contract DE-NA0003525. This paper, SAND2022-6270, describes objective technical results and analysis. Any subjective views or opinions that might be expressed in the paper do not necessarily represent the views of the U.S. Department of Energy or the United States Government.}}

\usepackage{amsopn}

\ifpdf
\hypersetup{
	pdftitle={\TheTitle},
	pdfauthor={\TheAuthors}
}
\fi

\begin{document}
	\maketitle
	
	\begin{abstract}
	The presence and evolution of defects that appear in the manufacturing process play a vital role in the failure mechanisms of engineering materials. In particular, the collective behavior of dislocation dynamics at the mesoscale leads to avalanche, strain bursts, intermittent energy spikes, and nonlocal interactions producing anomalous features across different time- and length-scales, directly affecting plasticity, void and crack nucleation. Discrete Dislocation Dynamics (DDD) simulations are often used at the meso-level, but the cost and complexity increase dramatically with simulation time. To further understand how the anomalous features propagate to the continuum, we develop a probabilistic model for dislocation motion constructed from the position statistics obtained from DDD simulations. We obtain the continuous limit of discrete dislocation dynamics through a Probability Density Function for the dislocation motion, and propose a nonlocal transport model for the PDF. We develop a machine-learning framework to learn the parameters of the nonlocal operator with a power-law kernel, connecting the anomalous nature of DDD to the origin of its corresponding nonlocal operator at the continuum, facilitating the integration of dislocation dynamics into multi-scale, long-time material failure simulations.
    \end{abstract}

\section{Introduction}

Dislocation dynamics is intrinsically connected to plasticity \cite{bulatov1998connecting} and material failure, emitted from crack tips \cite{zhu2004atomistic}, and piling-up leading to fatigue crack initiation \cite{tanaka_dislocation_1981}. The long-range interaction of dislocation stress-fields leads to collective motion characterized by avalanches, intermittency, and power-law scaling in energy and velocity distributions \cite{miguel_intermittent_2001,zaiser_scale_2006,hahner_fractal_1998}. Numerical simulations have successfully reproduced those features from discrete dislocation dynamics (DDD) models \cite{Arsenlis2007}. From a continuum perspective, early attempts of proposing evolution laws led to overly phenomenological models \cite{holt_dislocation_1970,walgraef1985dislocation}. Stochastic approaches have been proposed to account for uncertainties during dislocation motion \cite{hahner1996theory,kapetanou2015statistical}. Lately, continuum dislocation dynamics (CDD) emerged as another alternative for the continuous modeling of dislocation lines \cite{hochrainer2014continuum,hochrainer2015multipole}, yet still focused on explicit modeling of dislocation-dislocation interactions. A meaningful representation of the collective dynamics of dislocations that highlights the nonlocal, stochastic, and anomalous behavior of  dislocation ensembles in a fluid-limit continuous model is still missing. The use of nonlocal vector calculus for continuous modeling of dislocation dynamics is a natural, yet novel alternative.

Nonlocal models present an alternative to classic differential models where discontinuities are allowed, and long-range interactions are naturally present in an integral formulation. These features are attractive in the solution of problems involving convection-diffusion \cite{du2014nonlocal,d2017nonlocal}, heterogeneous media \cite{du2016multiscale}, turbulent flows \cite{akhavan2021data, seyedi2022data, SAZ2020fractional, akhavan2021nonlocal,SAZ2022tempered}, anomalous materials \cite{suzuki2021thermodynamically}, and subsurface dispersion \cite{schumer2001eulerian,xu2022machine}. For more applications, please refer to \cite{suzuki2021fractional} and references therein. The peridynamic theory \cite{silling_reformulation_2000} was proposed as a nonlocal alternative to classical continuum mechanics of solids, with applicability in fracture problems with discontinuities \cite{silling2003dynamic,silling_peridynamic_2014}. Over the last decade, a formalization of nonlocal models into a nonlocal vector calculus has been extensively discussed \cite{du2012analysis,du2013nonlocal}, along with advances towards the unification of nonlocal/fractional models \cite{d2020unified,d2013fractional}.

With the popularity of Machine Learning (ML) methods, several disciplines have seen increasing applicability of learning algorithms to enhance the understanding of the physics, to learn parameters of a model, or to construct robust surrogates based on high-fidelity data. Data-driven approaches for dislocation dynamics have lately acquired more interest. In \cite{salmenjoki2018machine}, authors used two-dimensional DDD simulations to train an algorithm for prediction of stress-strain curves. A ML approach for prediction of material properties from dislocation pile-ups was presented in \cite{sarvilahti2020machine}. Classification algorithms have also been used in the context of dislocation micro-structures \cite{salmenjoki2017predicting,steinberger2019machine}. Data-driven surrogate modeling of dislocation glide for computation of mobility estimates with uncertainty was proposed in \cite{de2021atomistic}. Other ML approaches have also grown in the context of learning the physics in the form of PDEs. We note the contributions of Physics-Informed Neural Networks (PINNs) \cite{raissi2019physics} which enhances deep neural networks with physics-based constraints, and PDE discovery approaches through the use of candidate terms and operators \cite{bakarji2021data,lee2020coarse,rudy2017data,supekar2021learning}.

The problem of learning kernels in integral operators has gained attention over the last years, with major contributions in the context of homogenization via nonlocal modeling and, more generally, in nonlocal and fractional diffusion diffusion. In one front, nPINNs \cite{pang2020npinns} was introduced as the nonlocal counterpart of the PINNs framework. Here, nonlocal equations are incorporated as constraints while training a deep neural network, for both forward and inverse problems with power-law kernel and finite horizon. The extraction of more complex kernels was investigated, via an operator regression approach, in \cite{you2021data}, allowing the possibility of sign-changing kernels by representing the kernel function through a polynomial expansion. This approach was further used in diverse applications such as peridynamics \cite{xu2021machine}, constitutive laws \cite{you2020data}, coarse-graining of molecular dynamics simulations \cite{you2021data2}, and homogenization of subsurface transport through heterogeneous media \cite{xu2022machine}.

In the present work, we use two-dimensional DDD simulations to generate probability distribution functions from shifted dislocation positions obtained from numerous realizations of the DDD problem. This approach gives us directly the Lagrangian dynamics of dislocation position. We transform the particle dynamics into a continuum Probability Density Function (PDF) evolving over time through an Adaptive Kernel Density Estimation method, generating a time-series of dislocation position PDFs. We propose a nonlocal model defined through a kernel-based integral operator for the evolution of the PDFs as the fluid-limit of the underlying stochastic process, and develop a ML framework to parameterize the nonlocal kernel, learning from the PDF snapshots generated from DDD data.

We summarize our main contributions below.

\begin{itemize}
	\item We obtain the probabilistic particle dynamics directly from DDD simulations. We highlight the effect of external loading and multiplication in the final probability distribution of dislocation position. Such differences are not evident in velocity distributions.
	\item We propose a general nonlocal equation to model the evolution of dislocation probability distributions in space, establishing the link between the discrete nature of dislocation dynamics at the mesoscale and the origin of its corresponding nonlocal operator at the continuum scale.
	\item We develop a ML framework to solve the inverse problem of recovering the parameters of the nonlocal equation from high-fidelity data. Specifically, we feed PDFs obtained from DDD simulations into the ML algorithm and obtain the parameters of the nonlocal power-law kernel, in terms of the fractional order $\alpha$, horizon $\delta$, and a linear coefficient. 
\end{itemize}

This work establishes, for the first time, a systematic, direct connection between the discrete anomalous dynamics of dislocations at the mesoscale to their ultimate effect in a continuum sense. With this mindset, we obtain a fast alternative to simulate dislocation dynamics through the nonlocal surrogate model, while still maintaining the underlying physics of micro-structural processes. This leads to a more efficient connection to macroscale problems such as visco-elasticity \cite{suzuki2021thermodynamically} and fracture \cite{barros2021integrated,de2021data}.

\smallskip
This paper is organized as follows: in Section \ref{sec:ddd}, we present the high-fidelity two-dimensional DDD simulation setting for single crystals under creep for three canonical conditions. In Section \ref{sec:data}, we describe the construction of dislocation PDFs from shifted positions and an Adaptive Kernel Density Estimation method. We introduce the nonlocal models in Section \ref{nlml_sec:nonlocal} and propose a nonlocal transport equation for the dislocation PDFs. We present a learning algorithm for the nonlocal kernel in Section \ref{sec:ml}. Then, in Section \ref{sec:results}, we report on the learning results based on a manufactured solution for three canonical DDD cases, along with discussions on the results. We present the conclusions in Section \ref{sec:conclusions}.

\section{Two-Dimensional Discrete Dislocation Dynamics}
\label{sec:ddd}

The simplified setup of a two-dimensional simulation, although lacking the curvature and natural multiplication mechanisms that a three-dimensional simulation provides, is still a robust and efficient way of observing the collective interactions of dislocation populations in a controlled manner. It allows the extraction of important quantities of interest, such as velocity distributions, stress and plastic strain evolution, and has been adopted in the literature to understand dislocation avalanches and power spectrum time-signals \cite{miguel_intermittent_2001,salmenjoki2018machine}. Therefore, here we adopt a two-dimensional discrete dislocation approach with the goal of learning the main characteristics of collective dislocation dynamics as a first step to translate such effects into a nonlocal continuum model. Particular implementation details will be explained in each section where necessary.

We consider a two-dimensional square domain of size $L$, populated with straight edge dislocations with directions along the $z$ direction, each of them with a Burgers direction $\bm{b} = \pm b$ along $x$, assuming a single-slip system. We assume there is no climb mechanism, so dislocations may only move in the $x$ direction.  

Immersed in an elastic continuum medium, dislocations create a long-range stress field, such that each dislocation is affected by the presence of all other dislocations in the crystal through an interaction stress, as well as any external stress $\sigma_{ext}$. Given the distance between dislocations $\bm{r}$, the dislocation-dislocation interaction stress, $\sigma_i(\bm{r})$, is given by \cite{anderson2017theory}

\begin{equation}
\sigma_i(\bm{r}) = \frac{\mu b}{2 \pi (1-\mu)} \frac{x(x^2 - y^2)}{(x^2 + y^2)^2},
\end{equation}

\noindent where $x$ and $y$ represent the distances between the edge dislocations in the $x$ and $y$ directions, respectively, $\mu$ is the shear modulus, and $\nu$ is the Poisson ratio. 

We simulate a domain in the bulk of the material, and assume it to be sufficiently far from any free surface, therefore Periodic Boundary Conditions (PBC) are needed. In order to apply the PBC and take into account all the long-range interactions, we include the forces due to infinite images of the simulation box. The exact form of the interaction stress is \cite{van1995discrete}

\begin{equation}
\sigma_i(\bm{r}) = \frac{\mu b}{2 (1-\mu)} \frac{1}{L} \frac{\sin(X) \left[\cosh(Y) - \cos(X) -Y\sinh(Y)\right]}{\left[\cosh(Y)-\cos(X)\right]^2},
\end{equation}

\noindent where $X = \frac{2 \pi x}{L}$ and $Y = \frac{2 \pi y}{L}$. 

Then, under the assumption that dislocation motion is overdamped under the \textit{viscous drag} regime, the equation of motion for the $i-$th dislocation along the $x$ direction, from the single-slip and no climb assumption, is

\begin{equation}
\frac{1}{M} \frac{d x_i}{dt} = b_i \left(\sum_{m\neq i}^{N} \sigma_i (\bm{r}_m - \bm{r}_i) + \sigma_{ext} \right),
\label{nlml_Eq:mobility}
\end{equation}

\noindent where $M$ is the dislocation mobility. All stress definitions refer to the shear stress component $\tau_{xy}$, such that the combination of $\sigma_{ext}$ and $\sigma_i$ result in the resolved shear stress acting on the dislocation. The resolved shear stress is the effective driver of motion in the edge dislocation.

We can solve this equation at first using a Forward-Euler scheme. For simplicity, we rescale the units and solve the problem with length in units of $b$, stress in units of $\sigma_0 = \frac{\mu}{2 \pi (1-\nu)}$, and time in units of $t_0 = \frac{1}{M \sigma_0}$.

The plastic strain resulting from dislocation motion can be computed following Orowan's relation

\begin{equation}
\gamma = \frac{1}{L^2} \sum_{i=1}^{N}b_i \Delta x_i.
\end{equation}

Beyond the constitutive relations that govern the dislocation glide velocity due to interactions and external stress, Eq.(\ref{nlml_Eq:mobility}), two-dimensional discrete dislocation dynamics simulations also need to consider other phenomenological aspects such as annihilation and multiplication.

Results from linear elasticity become invalid near the dislocation core due to nonlinearity in the stress field. Therefore, when two dislocations of opposite Burgers vector are within a distance $d_a$, they annihilated each other and are removed from the simulation.

Dislocation multiplication does not occur naturally in 2D simulations, as compared to 3D-DDD. In order to mimic the bowing of dislocation curves due to Frank-Read sources, we need to consider a phenomenological model. Here we follow the procedure in \cite{van1995discrete}, where we distribute $N_s$ dislocation sources randomly in the domain. At each time-step, we check at the sources the resulting shear stress, and compare it to a critical stress $\tau_c$. If the stress is above $\tau_c$ for more than $t_{nuc}$ time-steps, we generate a pair of dislocations with distance $L_{nuc}$, such that, in scaled units, 

\begin{equation}
L_{nuc} = \frac{1}{\tau_c}.
\end{equation}

\subsection{Representative Example: Single Crystal Under Creep}

We simulate three examples of single crystals under creep loading, following the setup from \cite{miguel_intermittent_2001}. We consider a domain of size $L = 300\ b$, initial number of dislocations of $N_0 = 1500$, and annihilation with a critical distance of $d_a = 2\ b$. We distribute 20 dislocation sources randomly throughout the domain, with mean critical nucleation time $\bar{t}_{nuc} = 10$, and critical distance with mean $\bar{L}_{nuc} = 50$, both parameters with variance of $10\%\ \bar{L}_{nuc}$. 

We test three representative examples to understand the effect of dislocation sources and external load in the form and parameterization of the final nonlocal kernel, where the external shear load is expressed using the definition of rescaled stress units, $\sigma_0 = \frac{\mu}{2 \pi (1-\nu)}$.

\begin{itemize}
	\item \textbf{Case 1:} load of $\sigma_{ext} = 0.0125\ \sigma_0$ without dislocation multiplication.
	\item \textbf{Case 2:} load of $\sigma_{ext} = 0.0125\ \sigma_0$ with dislocation multiplication.
	\item \textbf{Case 3:} load of $\sigma_{ext} = 0.0250\ \sigma_0$ with dislocation multiplication.
\end{itemize}

Case 1 represents locations inside the mechanical part without the presence of imperfections, impurities, or microcracks such that dislocations that are present inside the material do not multiply, and hence just glide until occasional annihilation. Therefore, Case 1 is representative of regions with lower internal stresses, less intense dislocation activity and plastic flow, and no evident rapid failure processes. 

Conversely, Cases 2 and 3 contain dislocation multiplication sources in a phenomenological way, representing regions in a component where we would normally observe higher degradation, under the presence of microcracks, voids, impurities and rough surfaces. Those characteristics are natural dislocation generators and are typically associated with failure regions. Therefore, in Cases 2 and 3 we are observing what happens near failure-inducing locations. 

The DDD simulations are executed with an in-house Python code running on Intel Xeon Gold 6148 CPUs with 2.40GHz. In all cases, we first let the system relax for 10000 time-steps of size $\Delta t = 1\ t_0$ with no external stress. This procedure leads to intense activity and annihilations until the dislocations reach a meta-stable configuration with about half the number of original dislocations. Then, we apply $\sigma_{ext}$ with a time-step of $\Delta t = 0.01\ t_0$ until final time of $T = 30000$ for Case 1, and $T = 25000$ for Cases 2 and 3.

\begin{figure}
	\centering
	\subfloat[Initial configuration.]{\includegraphics[width=0.4\textwidth]{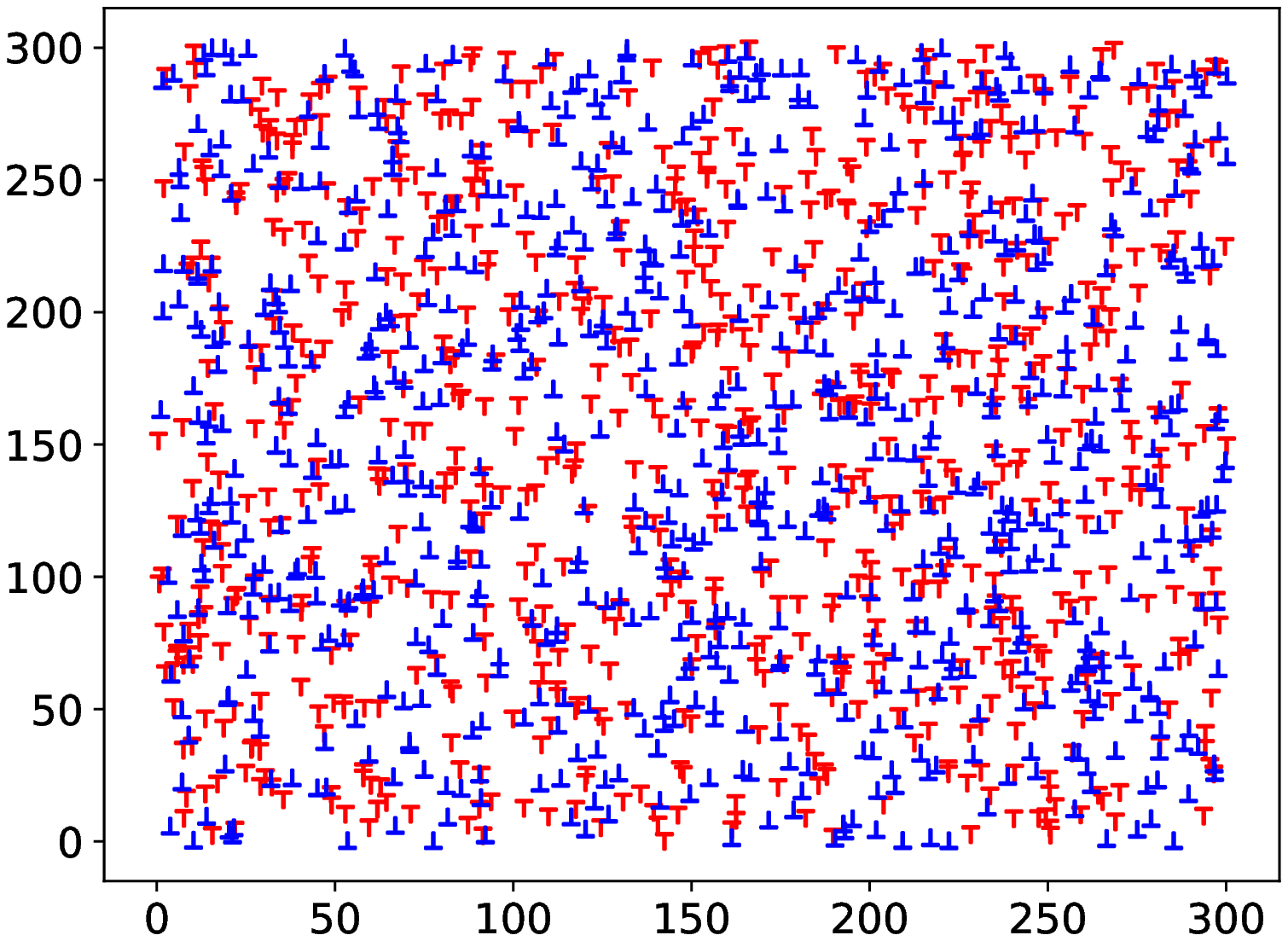}}
	\subfloat[Relaxed configuration.]{\includegraphics[width=0.4\textwidth]{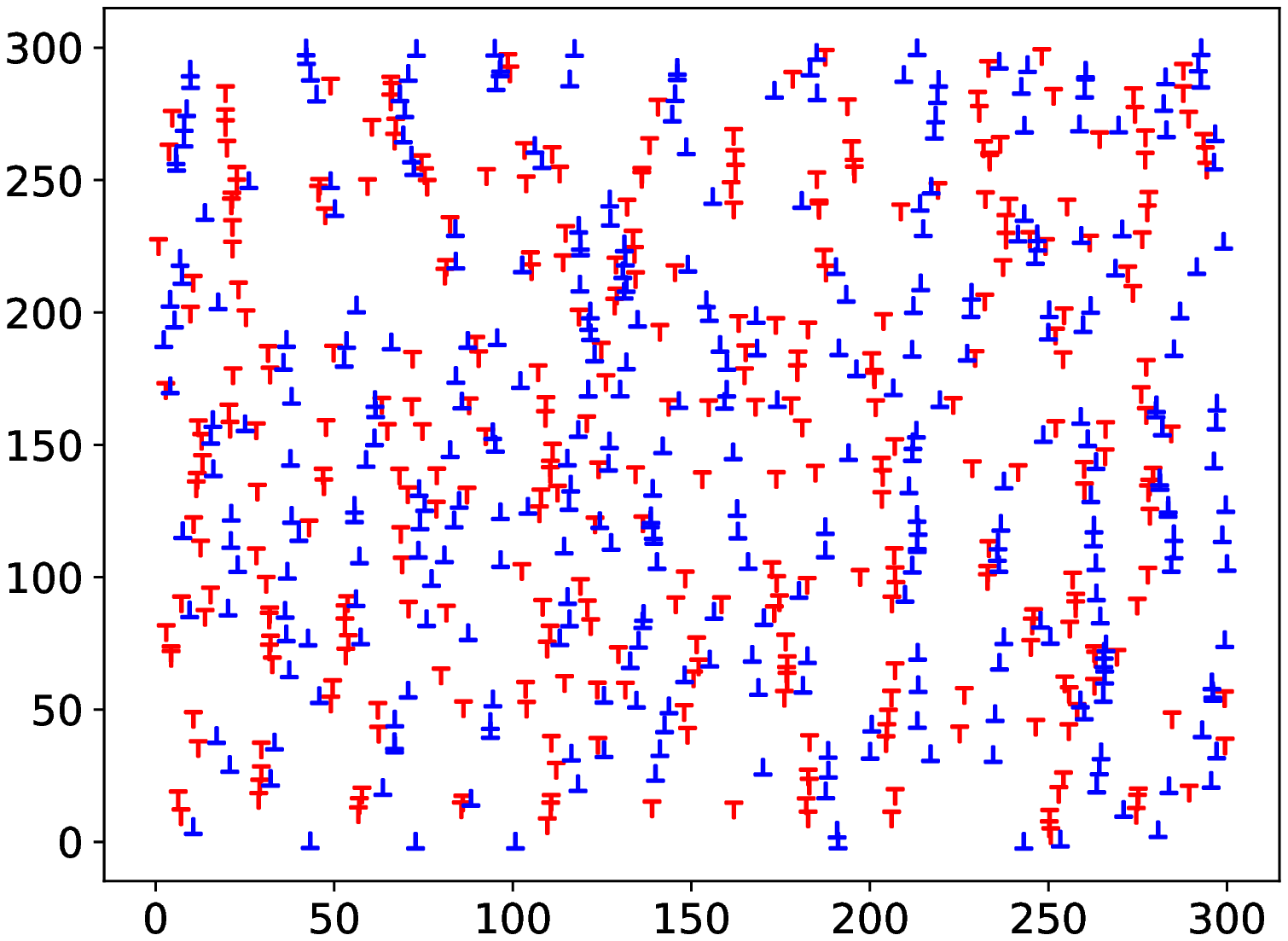}}
	\caption{Dislocation distribution at the beginning of the simulation (a), and after the relaxation (b) in a metastable structure for Case 1. Red and blue markers correspond to dislocations with positive and negative Burgers, respectively.}
	\label{nlml_fig:relax_config}
\end{figure}

\begin{figure}
	\centering
	\subfloat[Collective velocity.]{\includegraphics[width=0.33\textwidth]{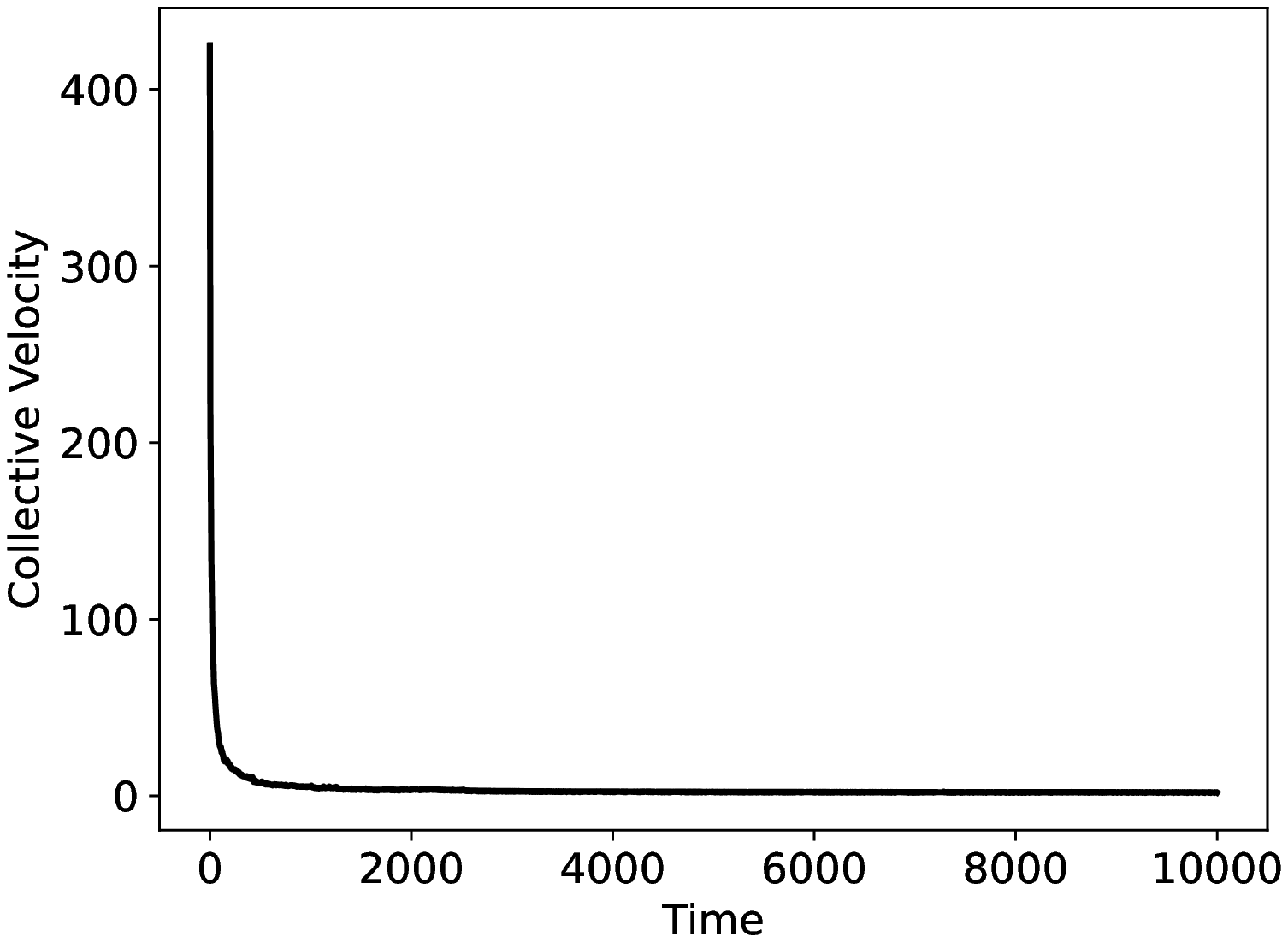}}
	\subfloat[Number of dislocations.]{\includegraphics[width=0.33\textwidth]{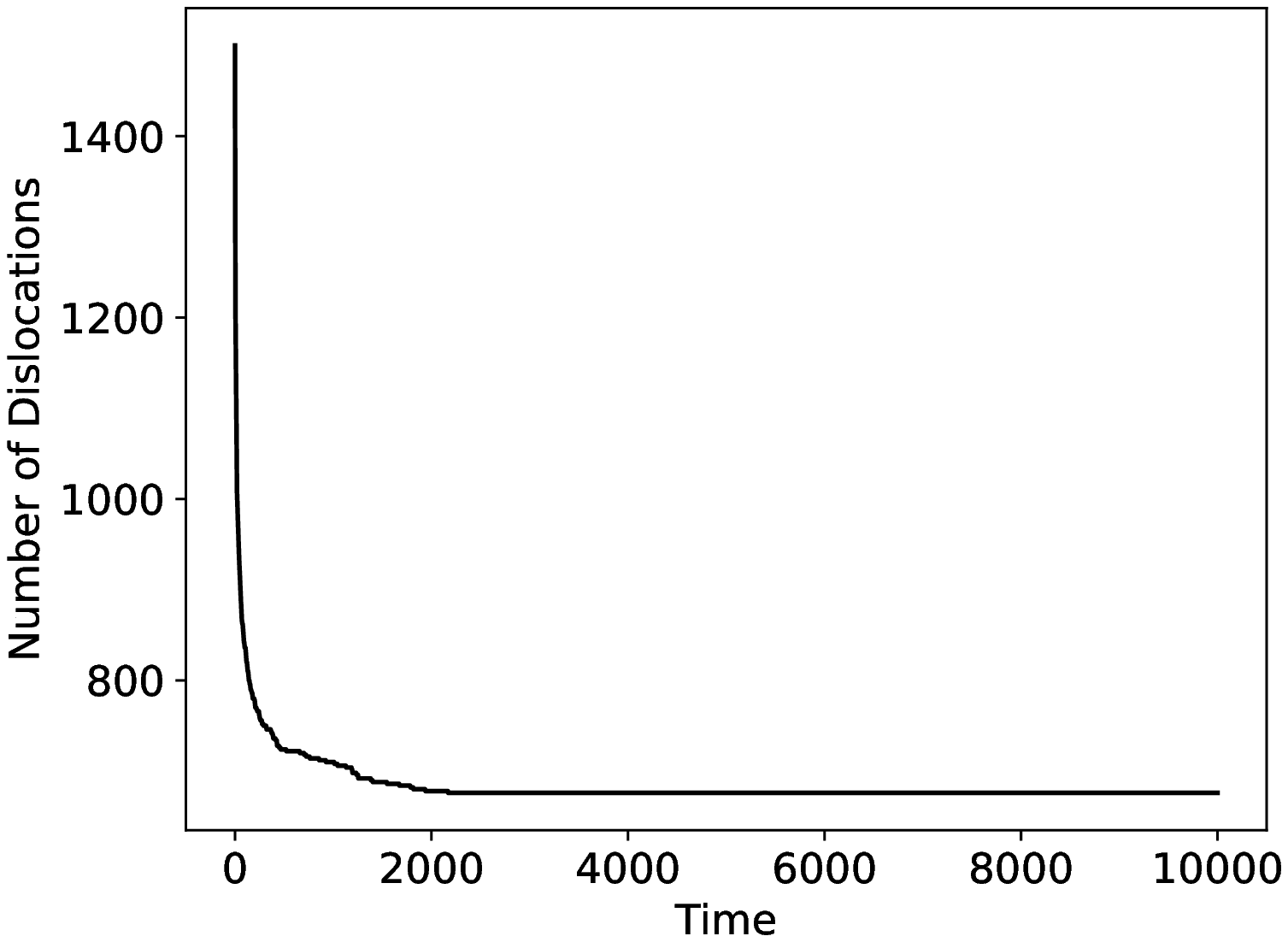}}
	\subfloat[Plastic strain $\gamma (t)$.]{\includegraphics[width=0.33\textwidth]{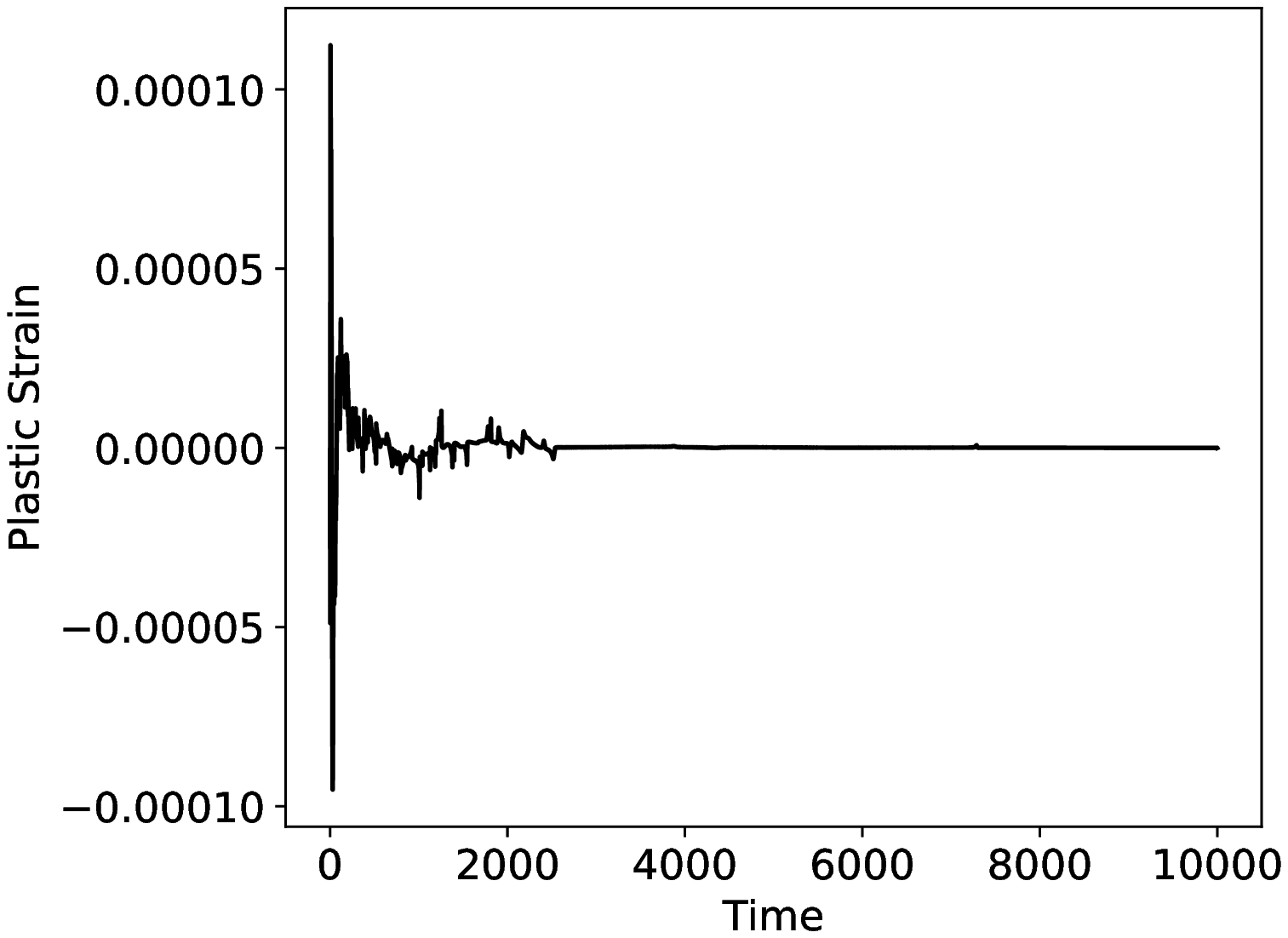}}
	\caption{Time-series plots of collective velocity, number of dislocations in the system, and plastic strain during the relaxation steps to show the system's stabilization. }
	\label{nlml_fig:relax_plots}
\end{figure}

Fig.~\ref{nlml_fig:relax_config} shows one realization of an initial dislocation configuration and the relaxed metastable configuration for Case 1. The time-series plots of the collective velocity $V = \frac{\sum_{i} v_i}{N}$, number of dislocations, and plastic strain during the relaxation steps are shown in Fig.~\ref{nlml_fig:relax_plots}.

\begin{figure}
	\centering
	\subfloat[Collective velocity.]{\includegraphics[width=0.33\textwidth]{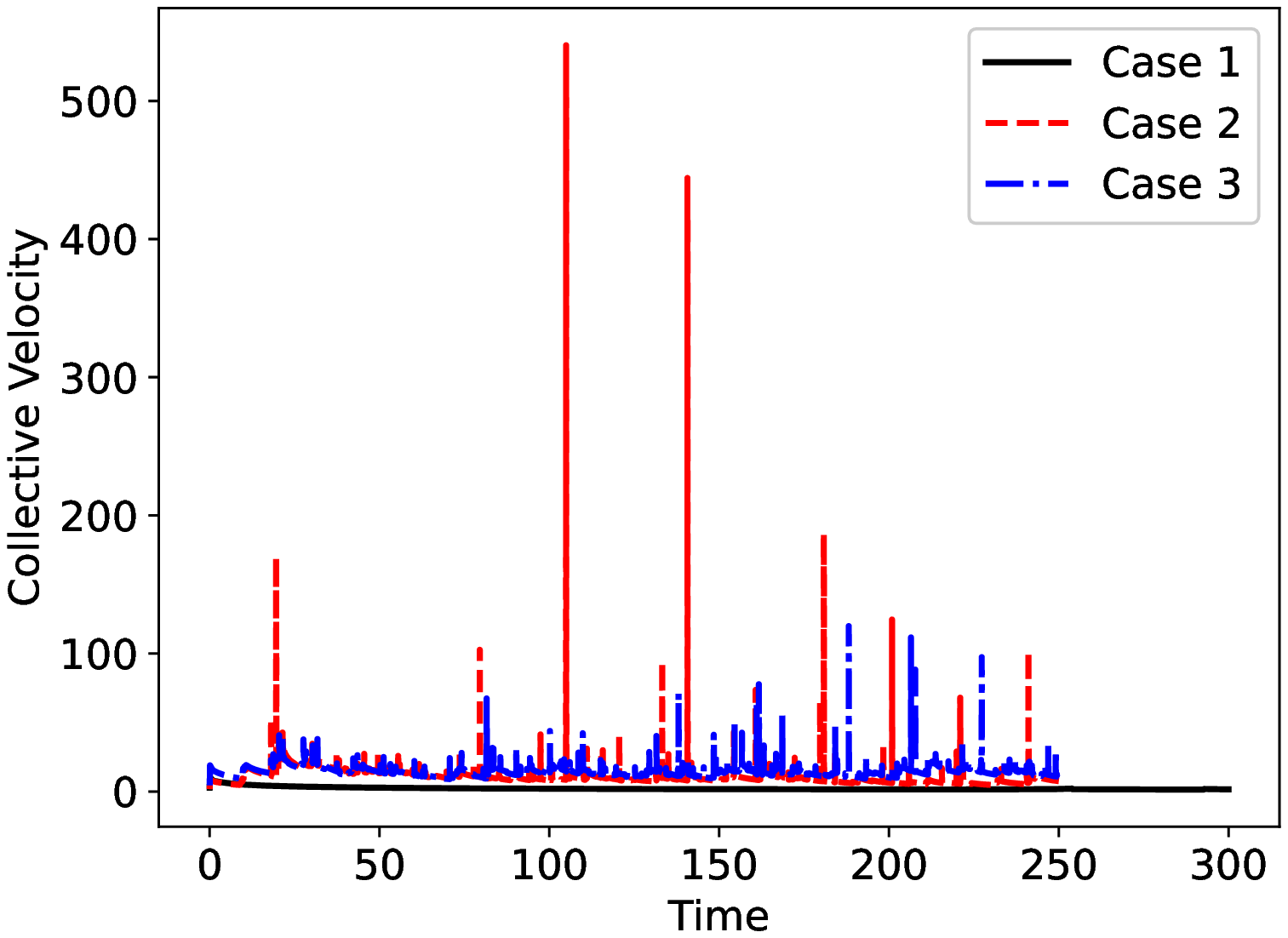}}
	\subfloat[Accumulated plastic strain.]{\includegraphics[width=0.33\textwidth]{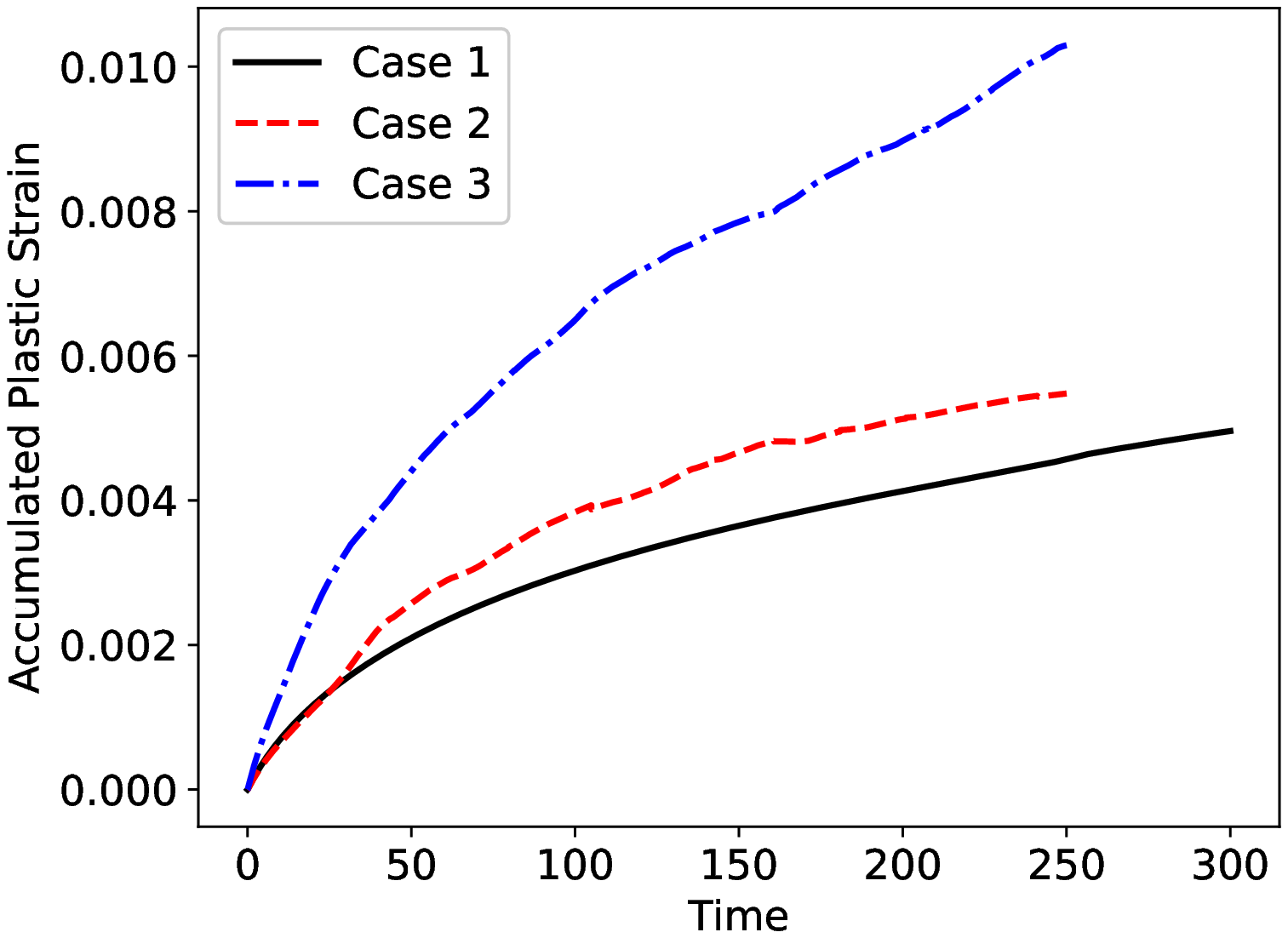}}
	\subfloat[Number of dislocations.]{\includegraphics[width=0.33\textwidth]{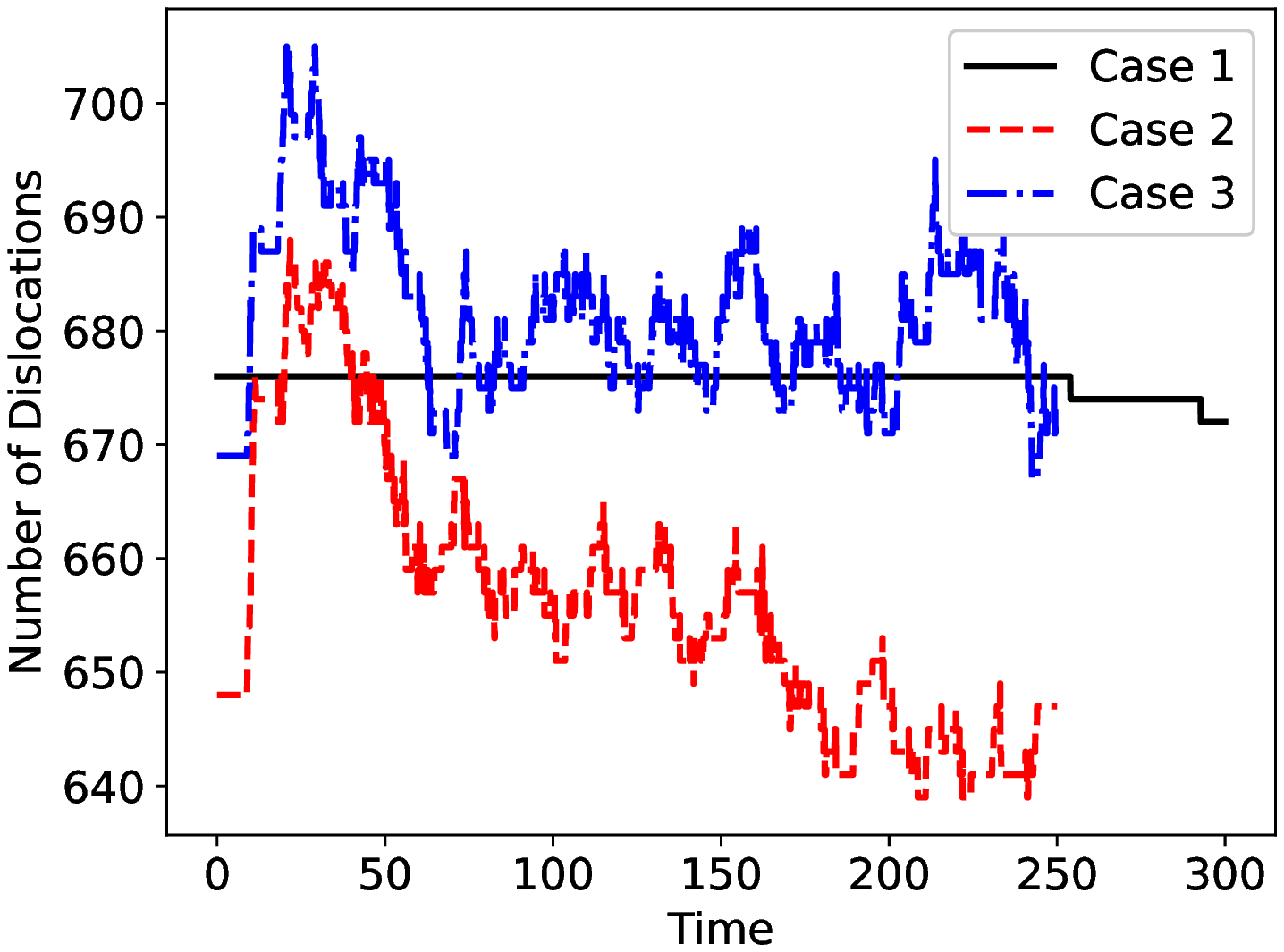}}
	\caption{Time-series plots of collective velocity, accumulated plastic strain, and number of dislocations in the system for a single realization of creep test for all three cases.}
	\label{nlml_fig:cases_ts}
\end{figure}

The collective statistics for the representative realizations discussed in this Section can be seen in Fig.~\ref{nlml_fig:cases_ts}. We plot the collective velocity, number of dislocations, and accumulated plastic strain during the creep load. The collective velocity signal is intermittent for Cases 2 and 3 with multiplication, where the spikes indicate bursts of activities during an avalanche. The higher load of Case 3 makes the baseline collective velocity be higher than Case 2, yet the spikes of Case 3 are not as large, since dislocations will tend to move faster, therefore having less time to interact close to the critical regime, as we see in Case 2. The higher baseline velocity also affects the accumulated plastic strain that is larger for Case 3. The number of dislocations is almost stable for Case 1, with Cases 2 and 3 showing more oscillations. This is due to the rearrangements that occur after a new dislocation pair is introduced, which eventually leads to more annihilations than Case 1.


\begin{figure}
	\centering
	\includegraphics[width=0.4\textwidth]{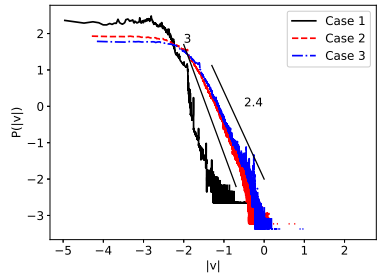}
	\caption{Probability Density Function of dislocation velocity for Cases 1, 2, and 3. We observe a power-law scaling of order $\beta = 2.4$ for Cases 2 and 3 with multiplication, and a sharper decay for Case 1.}
	\label{nlml_fig:dist}
\end{figure}

Last, we investigate the velocity statistics from the DDD simulations. Fig.~\ref{nlml_fig:dist} shows the PDF of individual dislocation velocity statistics collected throughout the whole simulation time for the single DDD realization of each case discussed in this section. We find that, in accordance with \cite{miguel_intermittent_2001}, the velocity PDFs show a power-law decay in the form $v \propto \sigma^{-\beta}$ with exponent around $\beta = 2.4$ for Cases 2 and 3 with multiplication. For Case 1, we see that the decay is slightly sharper. 

The velocity PDF has been extensively studied both experimentally and numerically over the past years, and strongly suggests that the nature of dislocation dynamics has anomalous characteristics. The power-law exponent of $\beta = 2.4$ is directly associated to intermittent velocity signals typical of avalanches and super-diffusive behavior. However, dislocation dynamics simulations, whether discrete or continuous are still too expensive to be run in the long time with the goal of understanding how the dislocation-network anomalous behavior influences the failure processes at the macroscale.

One could look at dislocation dynamics from the perspective of particle dynamics, where the dislocations would move under a underlying stochastic process. Ideally, we would analyze the statistics and construct a stochastic process that governs such particle dynamics, as a way of generating infinitely many particle trajectories that lead to the fluid-limit dynamics of the process. However, in the case of dislocation dynamics, even though the velocity distributions give us an idea of the type of stochastic behavior due to the heavy tails with power-law decay, the process cannot be simply described. Some dislocations are stuck, others jiggle around an equilibrium state, and a few others move intermittently with large velocity in a highly correlated motion.

Therefore, in order to obtain the fluid-limit dynamics, we still need to obtain statistics from a sufficient number of dislocation particles. The steps we take to this end will be discussed in the next section.

\section{Data Generation}
\label{sec:data}

In this section we describe the methodology to obtain empirical PDFs directly from DDD simulations, without the construction of a stochastic process, as discussed before. On the one hand, this makes the data generation process expensive, as it relies on simulation of multiple DDD simulations, instead of cheaper stochastic process trajectories. On other hand, this approach benefits from directly using high-fidelity dislocation data and is the most accurate representation of the dynamics we could generate.

\subsection{Obtaining Data of Shifted Positions}

We start by defining the shifted position $X_i(t)$ of a dislocation $i$ in a single realization of the DDD simulation. Given the initial position in the absolute frame of reference of the DDD box, $x_i(0)$, and the current absolute position in the simulation frame of reference, $x_i(t)$, the shifted position is a measure of the relative displacement of dislocation $i$ with respect to its initial position:

\begin{equation}
X_i(t) = x_i(t) - x_i(0),\quad \text{for} \quad t \in (0,T].
\end{equation}

We obtain a statistically significant collection of data-points $X_i(t)$ by considering the DDD simulation as stochastic. We run $n_r = 2000$ DDD simulations, each with random initial positions of dislocations and multiplication sources (for Cases 2 and 3). We distribute the execution of realizations among several HPC cores to take advantage of embarrassingly parallel stochastic simulations. With the number of dislocations after the relaxation between 600 and 700, the compilation of all dislocation shifted positions across the $n_r$ realizations gives the trajectories of $10^6$ Lagrangian particles that move following an underlying stochastic process starting at $X_i(0) = 0$. 

We take the Lagrangian particle trajectories obtained directly from DDD and translate them into an evolving PDF $\hat{p}(x,t)$, defined as the probability to find a dislocation at a distance $x$ from its initial position from the start of the DDD simulation. In the end, we want to construct a model for the time-evolution of $\hat{p}(x,t)$, and we propose that its evolution is governed by an integral operator that we model as a nonlocal Laplacian. In the following, we discuss how to transform the DDD position data into density estimates that will be used as training data for the ML algorithm.

\subsection{Density Estimation}

Under the proposed nonlocal model for evolution of dislocation position PDF, we are most interested in the nature of the dynamics of dislocation particles, i.e., how they react when put under creep stress along with an ever-changing stress landscape due to addition and removal of other dislocations. Given the focus on the dynamics due to load, multiplication, and annihilation mechanisms in a broader sense, in the limit of infinitely many particles, and not attempting to model the creation and destruction themselves, we do not include the birth/death in the nonlocal formulation. Instead, we only describe the nature of the dislocation motion from the continuum perspective, as a consequence of those mechanisms from the discrete representation at the DDD level.

\begin{figure}
 	\centering
 	\includegraphics[width=0.4\textwidth]{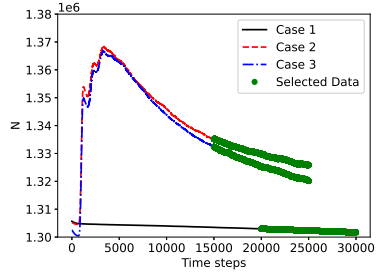}
 	\caption{Time-series of the total number of dislocations across the $n_r = 2000$ realizations of DDD for Cases 1, 2, and 3. We highlight the selected data for training and testing the ML algorithm.}
 	\label{nlml_fig:sel_data}
 \end{figure}

In this sense, the annihilation and creation of dislocations in DDD induce a level of noise when considering a continuous PDF representation. In the creep regime, we can minimize the interference of such noise by selecting a time domain over which the initial burst of dislocation multiplications and motion due to the rapid increase in stress up to the creep level has reached a steady-state regime. For training and testing of the ML framework, we select the last 10000 time-steps from the DDD simulations to generate the PDFs, as to minimize the effect of applying the creep load, and to obtain a data-set with the least changes in the number of particles. In the time-series selected, we observe only $0.1\%$, $0.69\%$, and $0.92\%$ in relative difference between final and initial number of dislocations in the selected data for Cases 1, 2, and 3, respectively. We can then assume a conservative nonlocal transport model in the continuum, yet annihilation, multiplication, and external load will directly influence the shape and parameterization of the nonlocal operator. We highlight the specific range of selected data over the whole time-series generated from DDD in Fig~\ref{nlml_fig:sel_data}. 

Before applying the estimator, we select the central $99.99\%$ of the probability mass, therefore discarding the outermost particles at each time-step. This procedure clearly defines a compact support for the PDF. 

We quickly summarize the classical and Adaptive Kernel Density Estimation formulations below.

\subsubsection{Kernel Density Estimation (KDE)} KDE are non-parametric estimators that do not require assumptions on the form of the sampling distribution. Yet, to use KDE, we need to apply a specific kernel  $k(x-x_i;w_0)$ parameterized by a bandwidth $w_0$. At this stage, we obtain an initial density estimate $\hat{p}_0(x)$ from

\begin{equation}
\hat{p}_0(x) = \frac{1}{n}\sum_{i=1}^n k(x-x_i;w_0),
\end{equation}

\noindent where $x$ is the coordinate over which we wish to evaluate the PDF, and $x_i$ are the positions of data points $i = 1,\,2,\,\dots,\,n$. 

Kernels are normalized to unity, i.e.,

\begin{equation}
\int_{-\infty}^{\infty}k(x;w_0)dx = 1,
\end{equation}

\noindent and take the form

\begin{equation}
k(x-x_i;w_0) = \frac{1}{w_0} K\left(\frac{x-x_i}{w_0}\right).
\end{equation} 

Then, the final density estimator is

\begin{equation}
\hat{p}_0(x) = \frac{1}{nw_0}\sum_{i=1}^n K\left(\frac{x-x_i}{w_0}\right).
\end{equation}

Let $s$ be the sample standard deviation. In the limit of number of data-points $n \to \infty$ of normally distributed data, the Mean Integrated Square Error of $\hat{p}_0$ is minimized when \cite{silverman2018density}

\begin{equation}
w_0 = 1.06 s n^{-1/5}.
\label{nlml_eq:w0}
\end{equation}

\subsubsection{Adaptive Kernel Density Estimation (AKDE)} Given the large jumps observed in DDD, we can expect the PDFs to have heavy tails, yet with limited data, while the majority of the mass would fall into the central part. A uniform binning method such as KDE would lead to the occurrence of noise in the tails, which we need to avoid as those curves feed the ML algorithm. Since higher density regions need narrower bins than low density tails, we use AKDE to obtain a continuous, smooth function.

We run the AKDE starting from an initial classical KDE estimate $\hat{p}_0(x)$ with a fixed bandwidth $w_0$ from Eq.~(\ref{nlml_eq:w0}). Then, the adaptivity takes place in the variable bandwidth for each data point \cite{silverman2018density}:

\begin{equation}
w_i = w_0 \lambda_i = w_0 \left( \frac{\hat{p}_0(X_i)}{G}\right)^{-\xi},
\end{equation}

\noindent where $G$ is defined as

\begin{equation}
G = \exp \left(\frac{1}{n} \sum_{i=1}^n \ln \hat{p}_0(X_i)\right).
\end{equation}

Furthermore, $0 \leq\xi\leq1$ is a sensitivity parameter that controls how important the shape of the initial guess is with respect to the second estimation \cite{pedretti2013automatic}. A theoretical optimal value was found to be $\xi = 0.5$ \cite{abramson1982bandwidth}. We finally obtain the density from AKDE as

\begin{equation}
\hat{p}_1(x) = \frac{1}{n} \sum_{i=1}^n \frac{1}{w_i} K\left(\frac{x-x_i}{w_i}\right).
\end{equation}

We define a domain $\Omega$ for the density function by adding a band of zeros beyond the right and left-most points in the $x$ direction, providing some space between the compact support of the PDF and the nonlocal simulation domain. We compute the estimates $\hat{p}_1$ at equally spaced points inside $\Omega$ defining a fixed grid size of $h = 0.05$. In Case 1, we have $\Omega_1 = [-15,15]$, and $m = 601$ points. Cases 2 and 3 are defined with $\Omega_2 = \Omega_3 = [-40,40]$ leading to $m = 1601$ points. Computation of $\hat{p}_1(x)$ for each time-step is also executed in parallel, each core processing a distinct time-step, for a total of 1000 HPC cores used for each case.

Finally, once the final estimates $\hat{p}_1(x)$ are obtained for all time-steps, in the last operation we enforce symmetry, as to avoid any inconsistencies in the ML algorithm, as we adopt a symmetric, radial-basis nonlocal kernel in the definition of our operator. We apply the symmetrization by checking the evolution of mean $\mu$ and skewness factor $\tilde{\mu}_3$ from $\hat{p}_1(x)$, defined as

\begin{equation}
\mu = \mathbb{E}[\hat{p}_1(x)] = \int \hat{p}_1(x) x  dx,
\end{equation}

\begin{equation}
\tilde{\mu}_3 = \mathbb{E}\left[\left(\frac{\hat{p}_1(x)- \mu}{\sigma}\right)^3\right] = \frac{\int (x-\mu)^3 \hat{p}_1(x) dx}{\left( \sqrt{\int (x-\mu)^2 \hat{p}_1(x) dx}\right)^3}.
\end{equation}

\begin{figure}
	\centering
	\subfloat[]{\includegraphics[width=0.25\textwidth]{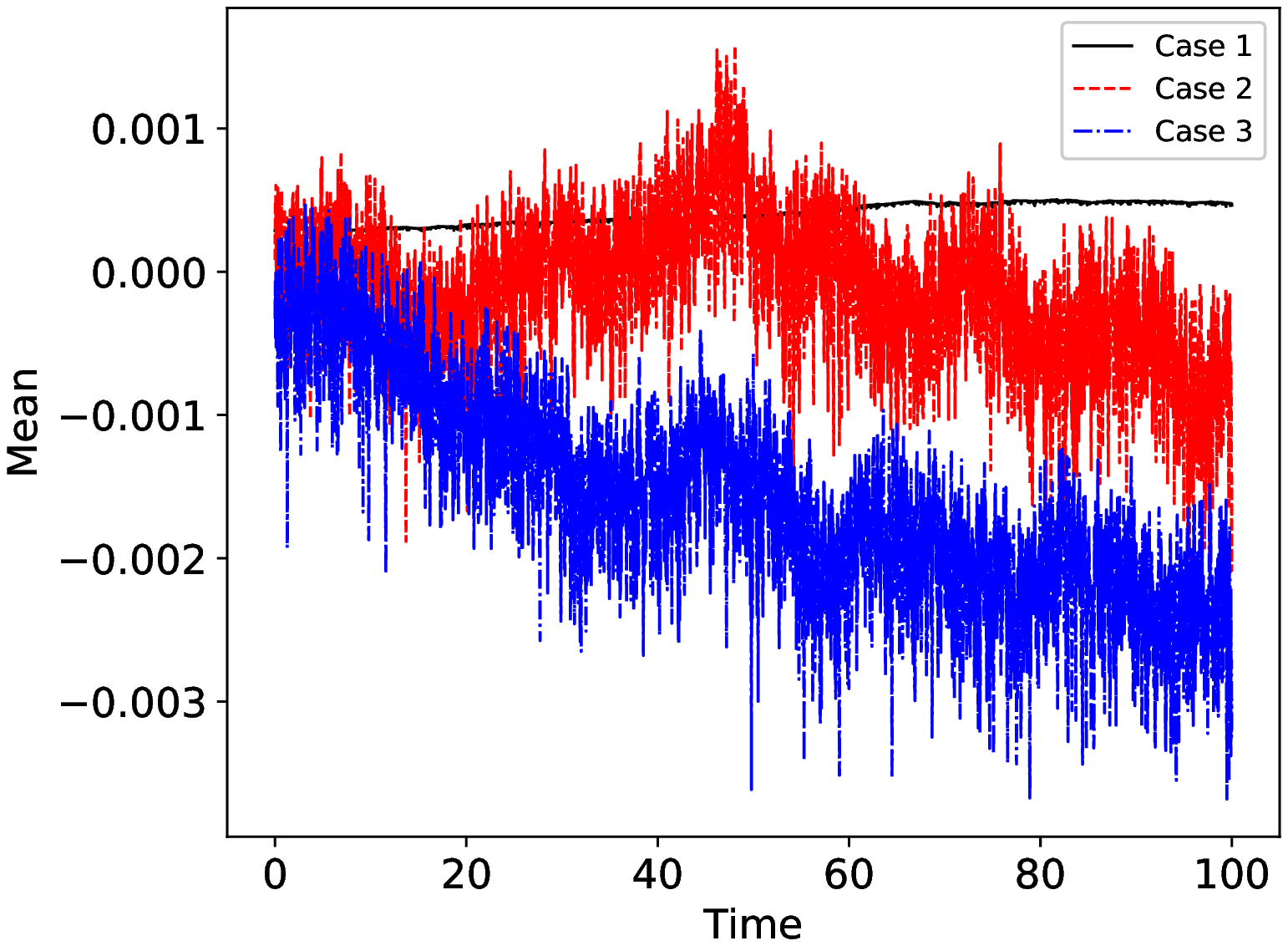}}
	\subfloat[]{\includegraphics[width=0.25\textwidth]{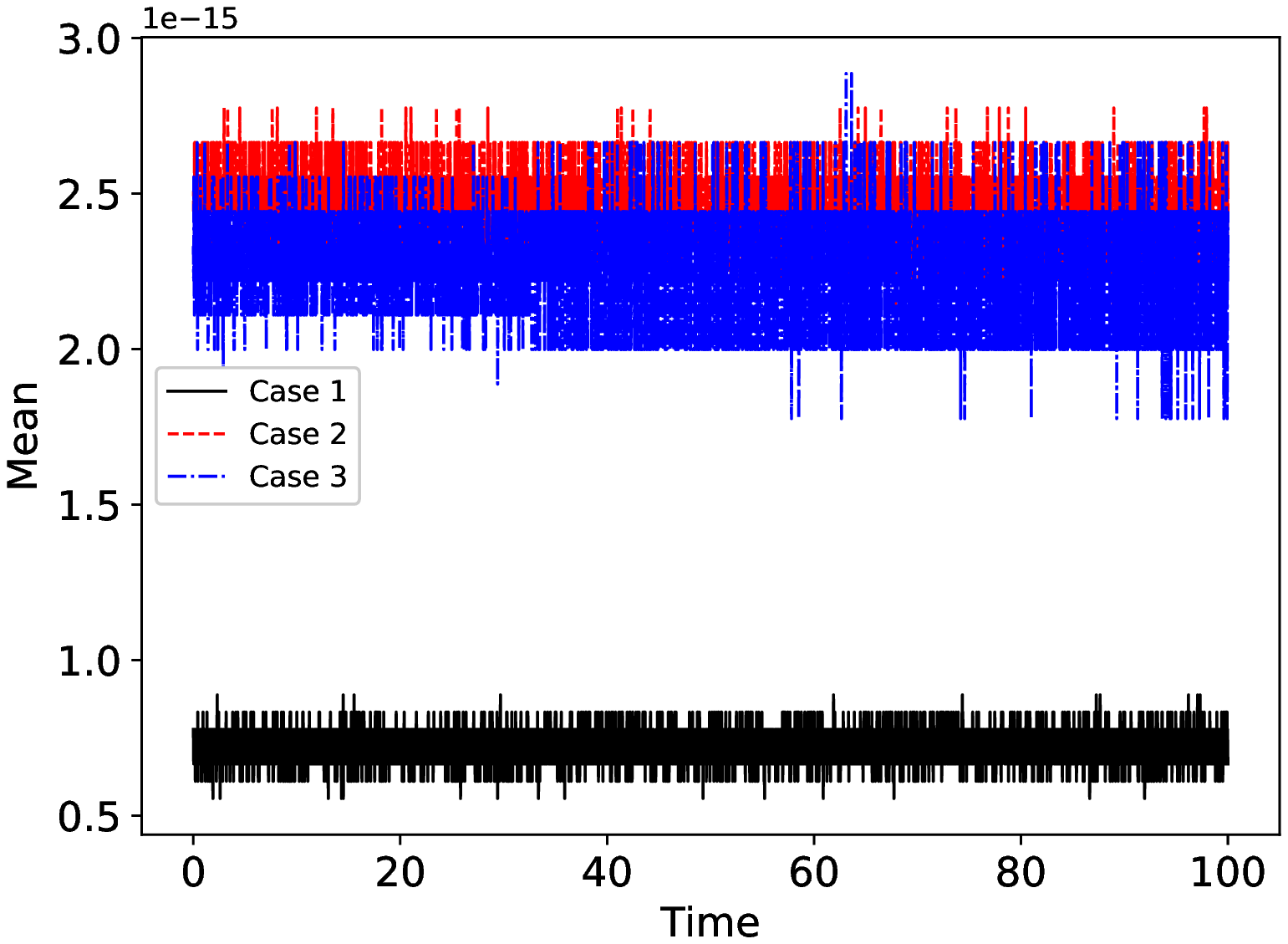}}
	\subfloat[]{\includegraphics[width=0.25\textwidth]{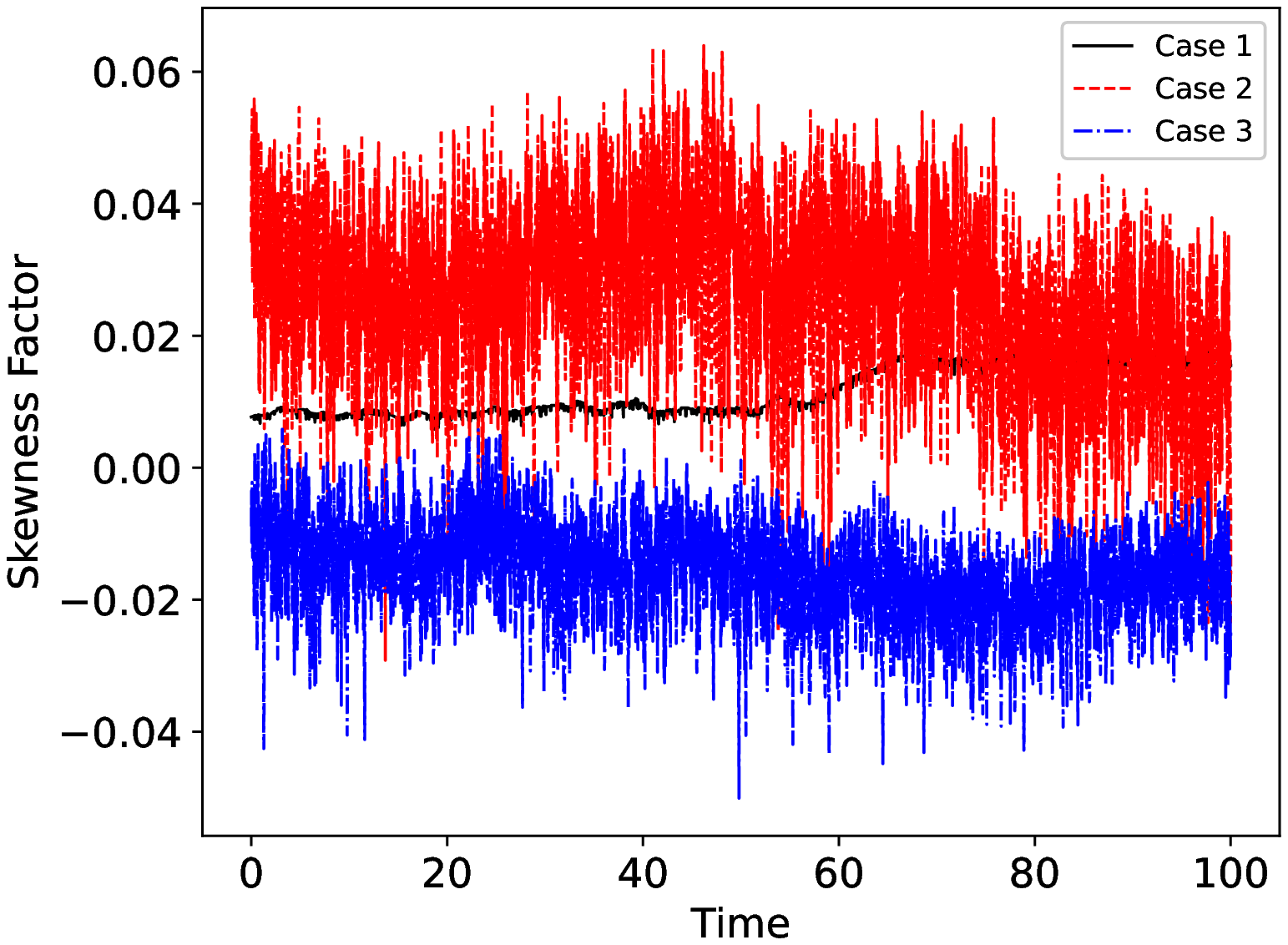}}
	\subfloat[]{\includegraphics[width=0.25\textwidth]{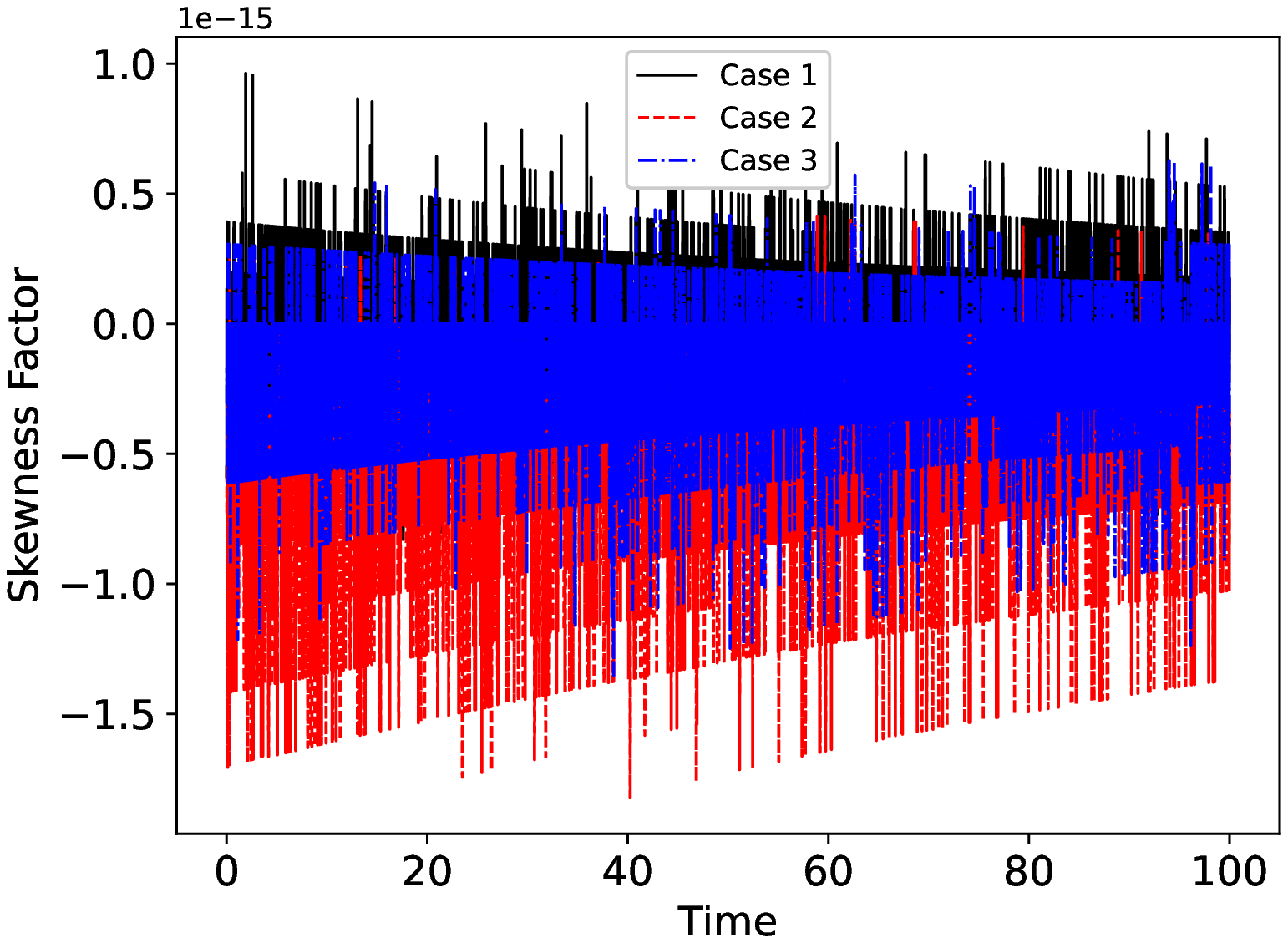}}
	\caption{Evolution of mean and skewness factor of $\hat{p}_1(x)$. The evolution of mean (a) before, and (b) after the symmetrization and re-normalization. The evolution of skewness factor (c) before, and (d) after symmetrization and re-normalization.}
	\label{nlml_fig:moments}
\end{figure}

We initially verify that they are sufficiently close to zero, assuming any non-zero measure of those parameters are due to lack of sufficient data. Then, we take the left side of $\hat{p}_1(x)$, mirror it to the other side, and re-normalize the PDF. We verify that after this procedure we guarantee a symmetric PDF that will better fit the radial-basis nonlocal kernel. Fig.~\ref{nlml_fig:moments} shows the values of $\mu(t)$ and  $\tilde{\mu}_3(t)$ before and after the symmetrization.

\begin{figure}
	\centering
	\subfloat[Case 1.]{\includegraphics[width=0.33\textwidth]{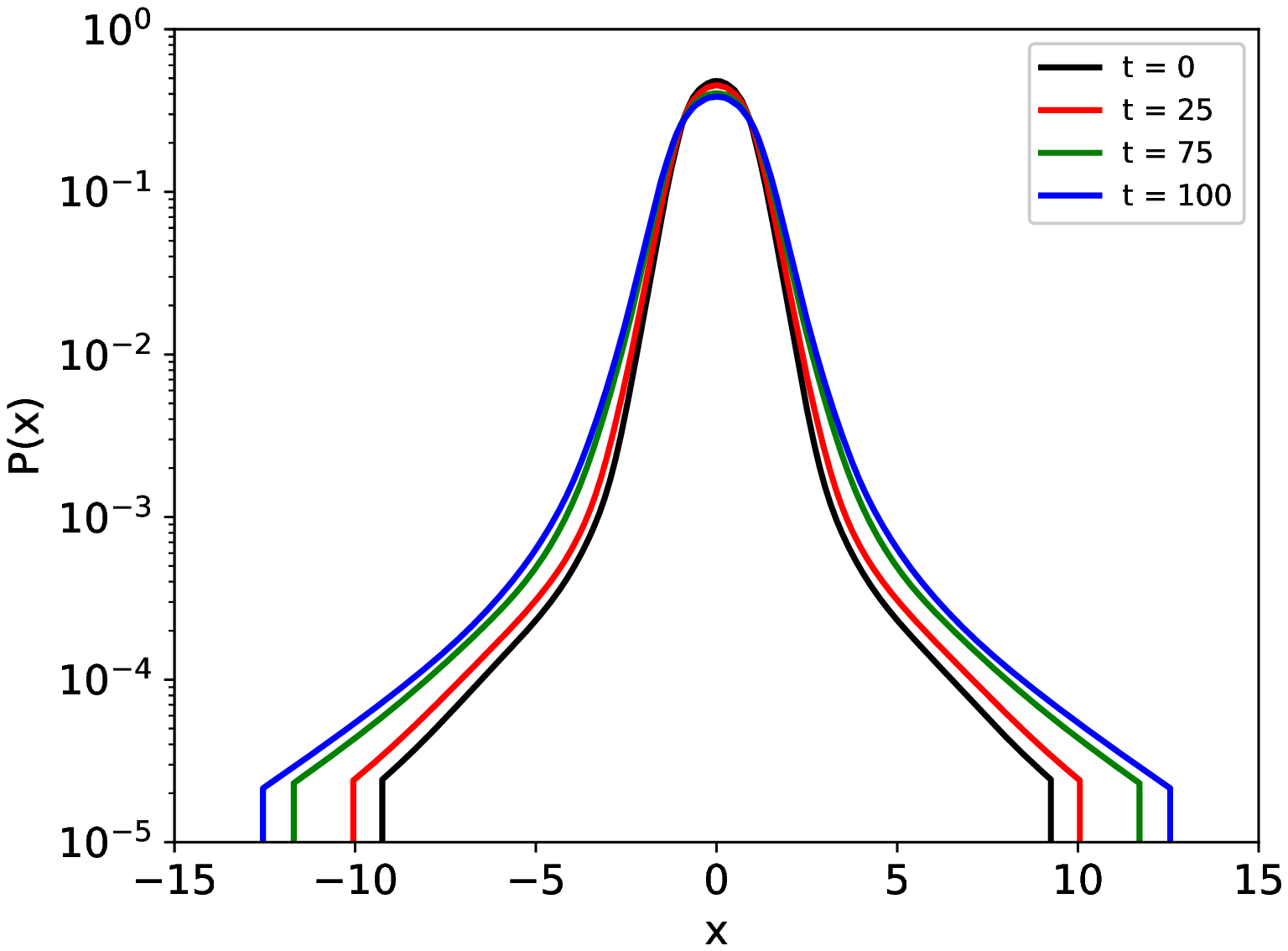}}
	\subfloat[Case 2.]{\includegraphics[width=0.33\textwidth]{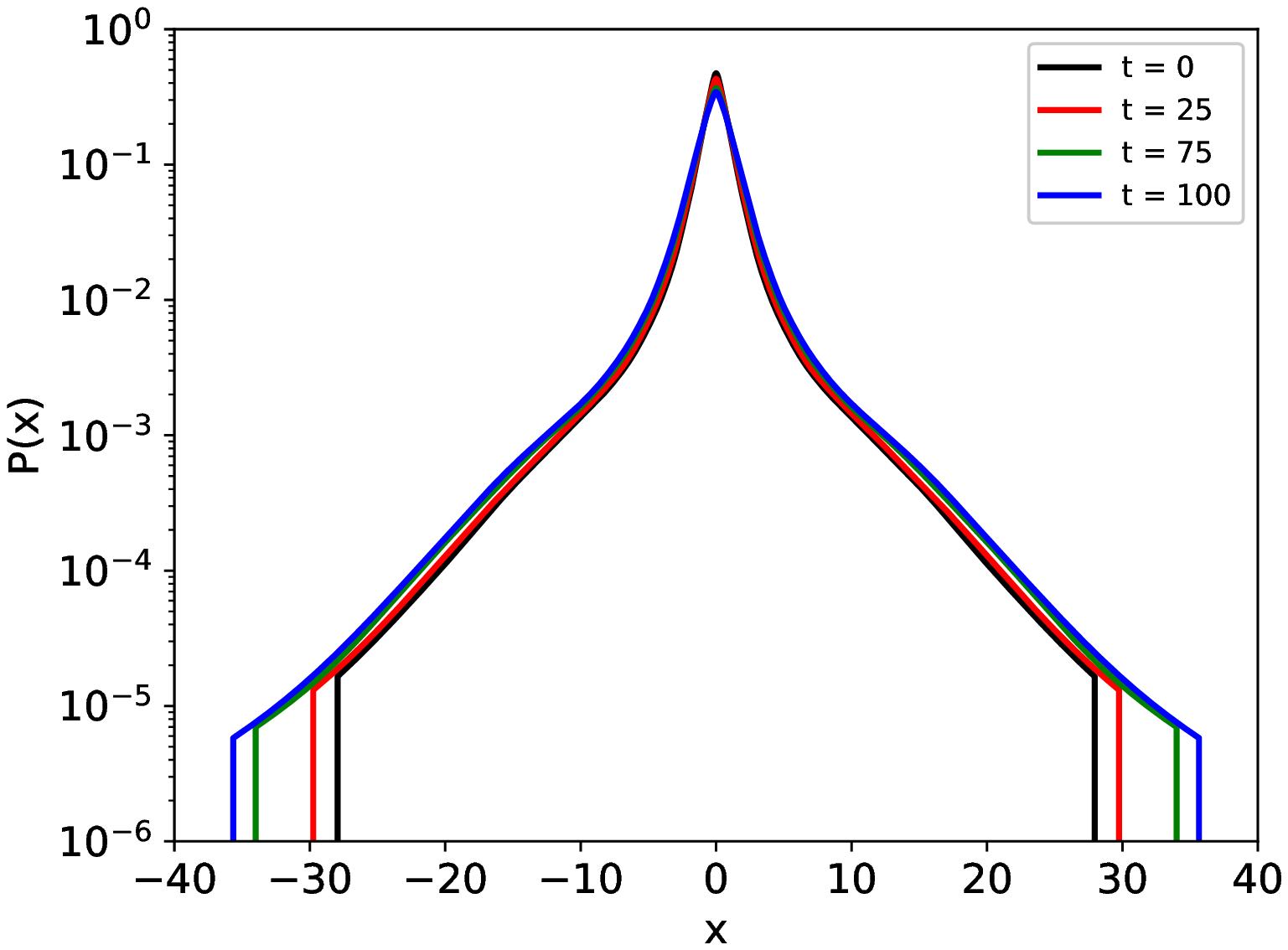}}
	\subfloat[Case 3.]{\includegraphics[width=0.33\textwidth]{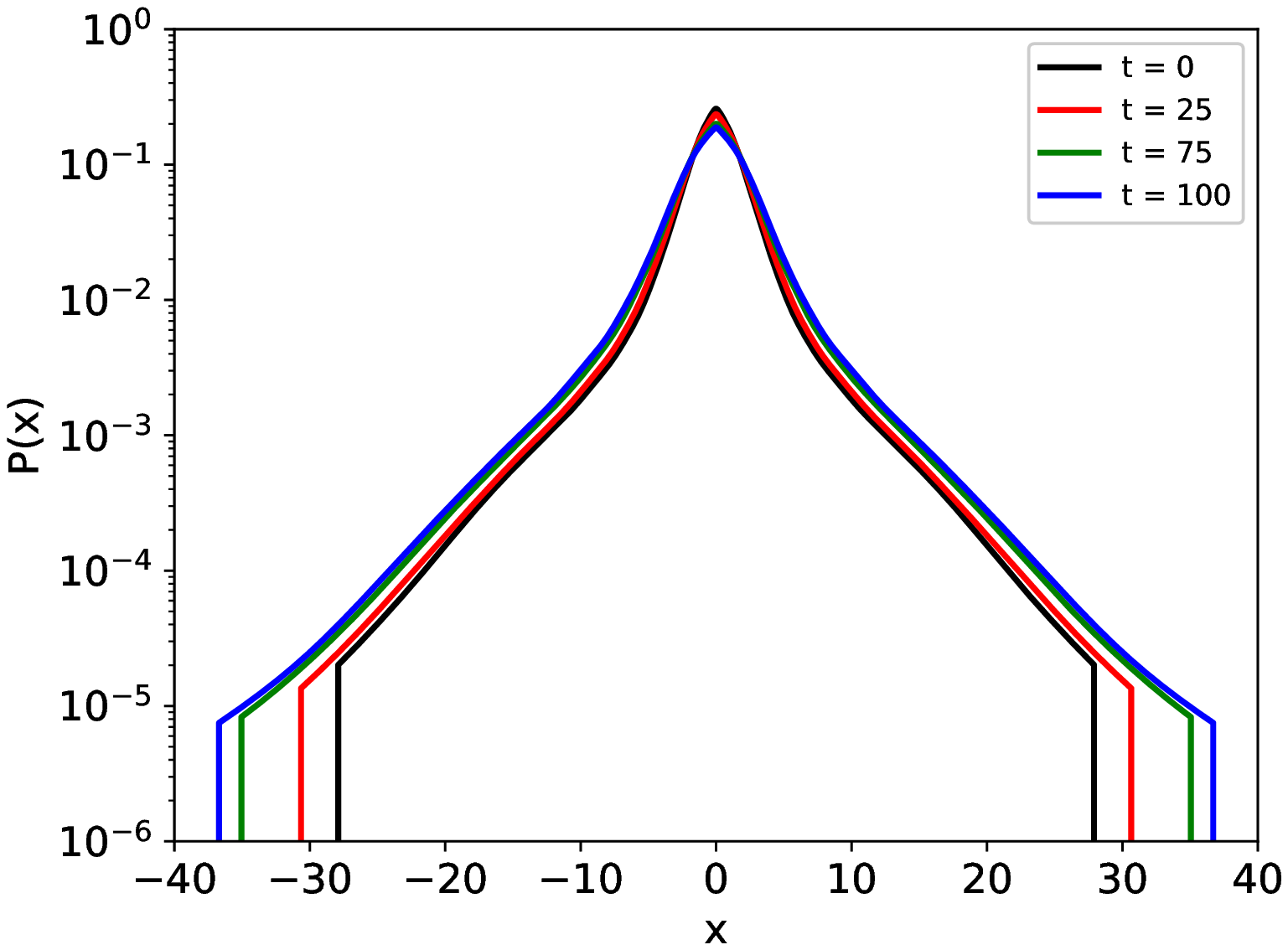}}
	\caption{Final shape of dislocation shifted position PDF from AKDE with symmetrization and re-normalization at selected time-steps.}
	\label{nlml_fig:pdfs}
\end{figure}

 Fig.~\ref{nlml_fig:pdfs} shows snapshots of the final PDF estimate for all three cases at selected time-steps. We plot the PDFs in logarithmic scale in the $y$ direction: we can readily see that the PDFs do not decay fast as one would expect from a Gaussian process. Instead, we have power-law decaying tails, with seemingly heavier ends on Cases 2 and 3 where multiplication mechanisms activate the collective dynamics more intensively. The heaviness on tails is accompanied by a larger support of the PDF. As we will see in the following sections, when we feed the PDFs into the learning algorithm, the kernel of the nonlocal operator will reflect those differences, establishing a meaningful link between the discrete and continuous dynamics.
 
\section{Nonlocal Transport Models}
\label{nlml_sec:nonlocal}

For the evolution of the dislocation position PDFs, we propose a parabolic nonlocal transport model defined through a nonlocal operator characterized by a kernel function, that we aim to determine. We use the data from DDD simulations to train a machine-learned surrogate model, for which we identify the model parameters.

We let $p(x,t)$ represent the empirical PDF estimate at time $t \in [0,T]$, and $x$ denote the position in over the domain $\Omega = [-L,L]$, where $L$ is defined by the taking the maximum support at the last time-step and including extra zeros, as seen in Fig.~\ref{nlml_fig:pdfs}~. We model the evolution of $p$ following the nonlocal parabolic equation

\begin{equation}
\begin{cases}
\dot{p}(x,t) = \mathcal{L}p,
\label{nlml_eq:nonlocal_evolution} & x \in \Omega\\
\mathcal{B}_{\mathcal{I}}p(x) = g(x), & x \in \Omega_\mathcal{I},
\end{cases}
\end{equation}

\noindent where $\mathcal{L}$ denotes the nonlocal (linear) Laplacian operator defined as

\begin{equation}
\mathcal{L}p = \int_{B_\delta(x)} K(\vert y - x \vert) (p(y) - p(x)) dy.
\label{nlml_eq:nonlocal_laplacian}
\end{equation}

$B_\delta(x)$ represents the ball centered at $x$ of radius $\delta$, also called the \textit{horizon}, defining the compact support of $\mathcal{L}$. It is relevant to note that for specific choices of kernel functions, $\mathcal L$ corresponds to well-known operators such as the fractional Laplacian \cite{d2020unified,d2013fractional}. In fact, when $K(|y-x|)\propto |y-x|^{-\alpha}$, with $\alpha=1+2s$, $s\in(0,1)$, the operator $\mathcal L$ corresponds to the one-dimensional fractional Laplacian. Furthermore, when the same kernel is restricted to the compact support $B_\delta(x)$, $\mathcal L$ corresponds to the so-called truncated fractional Laplacian. The latter turns out to be the operator of choice in our framework.

The interaction domain where nonlocal boundary conditions (or volume constraints) are prescribed is defined as:

\begin{equation}
\Omega_{\mathcal{I}} = \{ y \in \mathbb{R}\setminus\Omega \text{ such that } \vert y - x\vert < \delta \text{ for some }x \in \Omega\}.
\end{equation}

We prescribe nonlocal homogeneous Dirichlet volume constraints, given (in one dimension) by the nonlocal interaction operator $B_\mathcal{I}:[-L-
\delta,-L) \bigcup (L,L+\delta]\to\mathbb R$, such that $g(x) = 0$ at $x \in \Omega_\mathcal{I}$. 

The objective of the proposed ML algorithm is to train the nonlocal model on the basis of the series of PDFs; specifically, we find the best form and parameters of the kernel $K(\vert y - x \vert)$ such that Eqs.~(\ref{nlml_eq:nonlocal_evolution},\ref{nlml_eq:nonlocal_laplacian}) are satisfied and the predicted distributions are as close as possible to the high-fidelity dataset.
\section{Machine Learning of Nonlocal Kernels for Dislocation Dynamics}
\label{sec:ml}

\subsection{A Bi-level Machine Learning Framework}

We start the approximation by assuming that the kernel $K(\vert y - x \vert)$ is a radial function compactly supported on $B_\delta(x)$, decaying with $\alpha-$th order power-law, multiplied by a $P(\vert y - x \vert)$ function defined over $[0,\delta]$ 

\begin{equation}
K(\vert y - x \vert) = D\,\frac{ P(|y-x|)}{\vert y - x \vert^\alpha},
\label{nlml_eq:kernel}
\end{equation}

\noindent where we assume the coefficient $D \in \mathbb{R},\, D > 0$. The form of the function $P$ is part of the learning problem and its form strongly depends on the underlying physical system we want to reproduce. In the literature \cite{you2020data,you2021data,you2021data2}, the choice of a linear combination of Bernstein polynomials has been particularly successful. However, for the application considered in this work, the employment of Bernstein polynomials does not increase the surrogate's prediction power. For these reasons, we consider the simplified case of $P(|y-x|)=1$, for which the resulting operator corresponds to a truncated fractional Laplacian. Thus, the learning problem consists of finding the parameters $\alpha$, $\delta$, and the coefficient $D$ that parameterize the kernel.

We adopt a bi-level learning approach to reduce the dimensions of the minimization problem by exploiting the linearity of the nonlocal operator. Level 1 consists in obtaining the best coefficient $D$ for a given pair of parameters $\alpha$ and $\delta$, while at Level 2 the algorithm iterates over different values of $\alpha$ and $\delta$ and minimizes a cost function, each iteration using the optimal $D$ found in Level 1.

For the numerical solution of the bi-level optimization problem, we rewrite the nonlocal transport model, Eq.~(\ref{nlml_eq:nonlocal_evolution}), in a semi-discrete manner using a meshless approach, i.e.

\begin{equation}
\dot{p}(x_i,t) = \sum_{j \in \mathcal{H}} K(\vert x_j - x_i \vert) (p(x_j) - p(x_i)) h,
\end{equation}

\noindent where $H$ is the family of points $x_j$ in the neighborhood of point $x_i$, and $h$ is the distance between the points.

By using the power-law definition of the kernel in \eqref{nlml_eq:kernel}, with $P=1$, we can write the equation as

\begin{equation}
\dot{p}(x_i,t) = \sum_{j \in \mathcal{H}}  \frac{D}{\vert x_j - x_i \vert^\alpha} (p(x_j) - p(x_i)) h.
\label{nlml_eq:discrete}
\end{equation}

\subsubsection{Level 1}
We adapt the ideas presented in \cite{rudy2017data} for discovery of PDEs, yet, instead of identification of different PDE terms, our goal is to use the linear structure of Eq.~(\ref{nlml_eq:discrete}) to obtain the coefficient $D$ given a specific pair of values $(\alpha,\delta)$.

For given values of $\delta$ and $\alpha$, we construct vectors $U$, and $U_t$. $U$ contains the RHS of Eq.~(\ref{nlml_eq:discrete}), where the spatio-temporal data are reshaped into a single stacked column array. $U_t$ is the LHS of Eq.~(\ref{nlml_eq:discrete}) with the time-derivative computed through a forward Euler method at all space and time points, also transformed into a single column array. Given $n$ time-steps and $m$ grid points, both $U$ and $U_t$ have size $nm$.

Our problem $U_t = U D$ reads:

\begin{equation}
\begin{bmatrix}
\dot{p}(x_0,t_0)\\\dot{p}(x_1,t_0)\\\dot{p}(x_2,t_0)\\\vdots\\
\dot{p}(x_{m-1},t_n)\\\dot{p}(x_m,t_n)
\end{bmatrix} = D
\begin{bmatrix}
C (x_0,t_0) \\
C (x_1,t_0)  \\
C (x_2,t_0)  \\
\vdots \\
C (x_{m-1},t_n)\\
C (x_{m},t_n) 
\end{bmatrix}
\label{nlml_eq:matrix}
\end{equation}

\noindent with

\begin{equation}
C (x_i,t_k) = \sum_{j \in \mathcal{H}} \frac{1}{\vert x_j(t_k) - x_i(t_k) \vert^\alpha} (p(x_j(t_k)) - p(x_i(t_k)) h.
\end{equation}

Then, for every pair of $\alpha$, $\delta$ being considered in the minimization, we use a Least Squares solver to obtain the best $D$.

\subsubsection{Level 2}

We use a minimization algorithm to find $\alpha$ and $\delta$ that minimize the Mean Logarithmic Absolute Error (MLAE):

\begin{equation}
\epsilon =  \frac{1}{nm} \sum_{l=1}^{nm} \vert \log(p_l) - \log(\tilde{p}_l) \vert,
\label{nlml_eq:error}
\end{equation}

\noindent where $p_l$ represents the true value of the function at a particular $x$ ant $t$, and $\tilde{p}_l$ is the solution of the nonlocal model at $x$ and $t$ starting from the initial conditions at $t = 0$, for the current trial values $\alpha_{\text{trial}}$ and $\delta_{\text{trial}}$, and its corresponding $D$. We adopt the MLAE with the goal of giving as much significance to the information on the tails as we give to the central part of the PDF.

For all time-steps, we obtain $\tilde{p}$ at time-step $k+1$ and grid-point $i$, $\tilde{p}_i^{k+1}$, from  $\tilde{p}_i^{k}$, using the forward Euler scheme. Thus, Eq.~\eqref{nlml_eq:discrete} becomes

\begin{equation}
\tilde{p}_i^{k+1} = \tilde{p}_i^k + \Delta t\sum_{j \in \mathcal{H}}  \frac{D}{\vert x_j - x_i \vert^\alpha} (\tilde{p}^k_j - \tilde{p}^k_i) h.
\label{nlml_eq:full_discrete}
\end{equation}

For the solution of the minimization problem we adopt, among several possible choices, the Nelder-Mead Method (NM), which is a gradient-free, downhill simplex approach that uses a direct search method (based on function comparison). The overall bi-level algorithm for the identification of kernel parameters based on minimization by NM is presented in Algorithm~\ref{nlml_algo:NM}.

\begin{algorithm}[t]
	\caption{Bi-Level Machine Learning Algorithm with Nelder-Mead Minimization.}
	\label{nlml_algo:NM}
	\begin{algorithmic}[1]
		\State Choose the initial guess $(\alpha_0,\sigma_0)$.
		\For{Each iteration $i$ of NM (Level 2)}
		\State Construct the matrix equation Eq.~(\ref{nlml_eq:matrix}) and obtain the coefficient $D$ with trial parameters $(\alpha_i,\sigma_i)$ (Level 1).
		\State Solve the nonlocal model and obtain the trial solutions $\tilde{p}$ using Eq.~(\ref{nlml_eq:full_discrete}).
		\State Using the true $p$ and trial solutions $\tilde{p}$, Compute the error using Eq.~(\ref{nlml_eq:error}).
		\EndFor
		\State The algorithm gives the optimal parameters $(\alpha_{\text{opt}},\sigma_{\text{opt}})$ and associated $D_{\text{opt}}$.
	\end{algorithmic}
\end{algorithm}

In the solution of the inverse problem, the advantages of high-performance computing become more evident, since we solve a regression problem and simulate 10000 time-steps of a nonlocal diffusion equation at each iteration. Therefore, it is paramount that we exploit parallelism in the solution of Algorithm~\ref{nlml_algo:NM}. We implement the learning algorithm in Python; we make use of the NumPy library for the Least Squares regression, and SciPy for the minimization, using the built-in Nelder-Mead method. We parallelize both Level 1 and 2 using the MPI4Py library. At Level 1, each processor computes a section of the RHS, as they are independent computations from the already available training dataset. The LHS is computed once at the beginning of the algorithm, as it is constant for all iterations. Then, in Level 2, at each time-step of Eq.~(\ref{nlml_eq:full_discrete}), we parallelize the computation of $\tilde{p}_i^{k+1}$. In the end, the parallel implementation speeds up the costly computation of nonlocal operators, and allows the algorithm to converge in less than two hours for Case 1, with $m = 601$ grid points running 200 cores, and in slightly more than two hours for Cases 2 and 3, with $m = 1601$ grid points using 400 cores.

\section{Results and Discussion}
\label{sec:results}

\subsection{Method of Manufactured Solution}

We assess the learning algorithm and nonlocal modeling proposed via the Method of Manufactured Solutions, where we produce training data from a known kernel and recover it through the ML algorithm as a necessary consistency check of the proposed ML framework. Starting from an initial condition, we solve the nonlocal diffusion equation, Eq.~(\ref{nlml_eq:nonlocal_evolution}), with the kernel parameterized by known  $\alpha_{\text{true}}$, $\delta_{\text{true}}$, and $D_{\text{true}}$, generating the snapshots of $p(x,t)$ to be provided to the ML algorithm.

We simulate the nonlocal diffusion problem, Eq.~(\ref{nlml_eq:discrete}), in a domain $\mathcal{D} = [-1,1]$ with $L = 2$, and select $\alpha_{\text{true}} = 1.5$, $\delta_{\text{true}} = L/2$, and $D_{\text{true}} = 0.1$ as our parameters. For comparison, we simulate the nonlocal diffusion problem with spatial discretization using $m = 101$ points in space, solving the equation over $n = 200$ time-steps in size with $\Delta t = 0.01$. The initial condition for the nonlocal diffusion problem is a Dirac delta function at $x = 0$ with area equal to 1. Similarly to the DDD dataset, we let the system evolve and only use the last 200 time-steps of the simulation to collect the training and testing sets, using $80\%$ of time-steps for training, and the rest for testing.

We compare the relative errors of $\alpha$, $\delta$, and $D$ by the following expression

\begin{equation}
\varepsilon_i = \frac{\vert \xi_{i,\text{opt}} - \xi_{i,\text{true}} \vert }{\vert \xi_{i,\text{true} }\vert },
\end{equation}

\noindent where $\xi_i$ represents the parameter in consideration,  \textrm{opt} corresponds to the optimal value found by the algorithm, and \textrm{true} denotes the true parameter value.

We adopt a parametric space with bounds of $\alpha = [0,4]$, and $\delta = [\Delta x,L]$. Given the true values of  $\alpha_{\text{true}} = 1.5$ and $\delta_{\text{true}} = L/2$, we take the initial guess to be  $\alpha_{0} = 2$, $\delta_{0} = L/4$. We present the parameter, training and testing error results in Table~\ref{nlml_tab:manuf}. We verify that the algorithm successfully identifies the parameters within a maximum of $1\%$ error for the horizon $\delta$, while $\alpha$ and $D$ are within $0.01\%$ error. This example showcases that the decoupling of the learning in 2 levels leads to the correct kernel. Given the higher number of training points, and the time-dependent dynamics, it is expected to have a lower testing error, compared to training.

\begin{table}[]
	\centering
	\caption{Parameter and algorithm errors for the manufactured solution}
	\label{nlml_tab:manuf}
	\begin{tabular}{lllll}
		\toprule
		$\alpha$  & $\delta$ & $D$ & Training & Testing \\ \midrule
		9.81e-4        & 1.02e-2        & 6.63e-4   & 9.50e-4        & 1.06e-4      \\ \bottomrule
	\end{tabular}
\end{table}

\begin{figure}[]
	\centering
	\subfloat[Iteration errors.]{\includegraphics[width=0.4\textwidth]{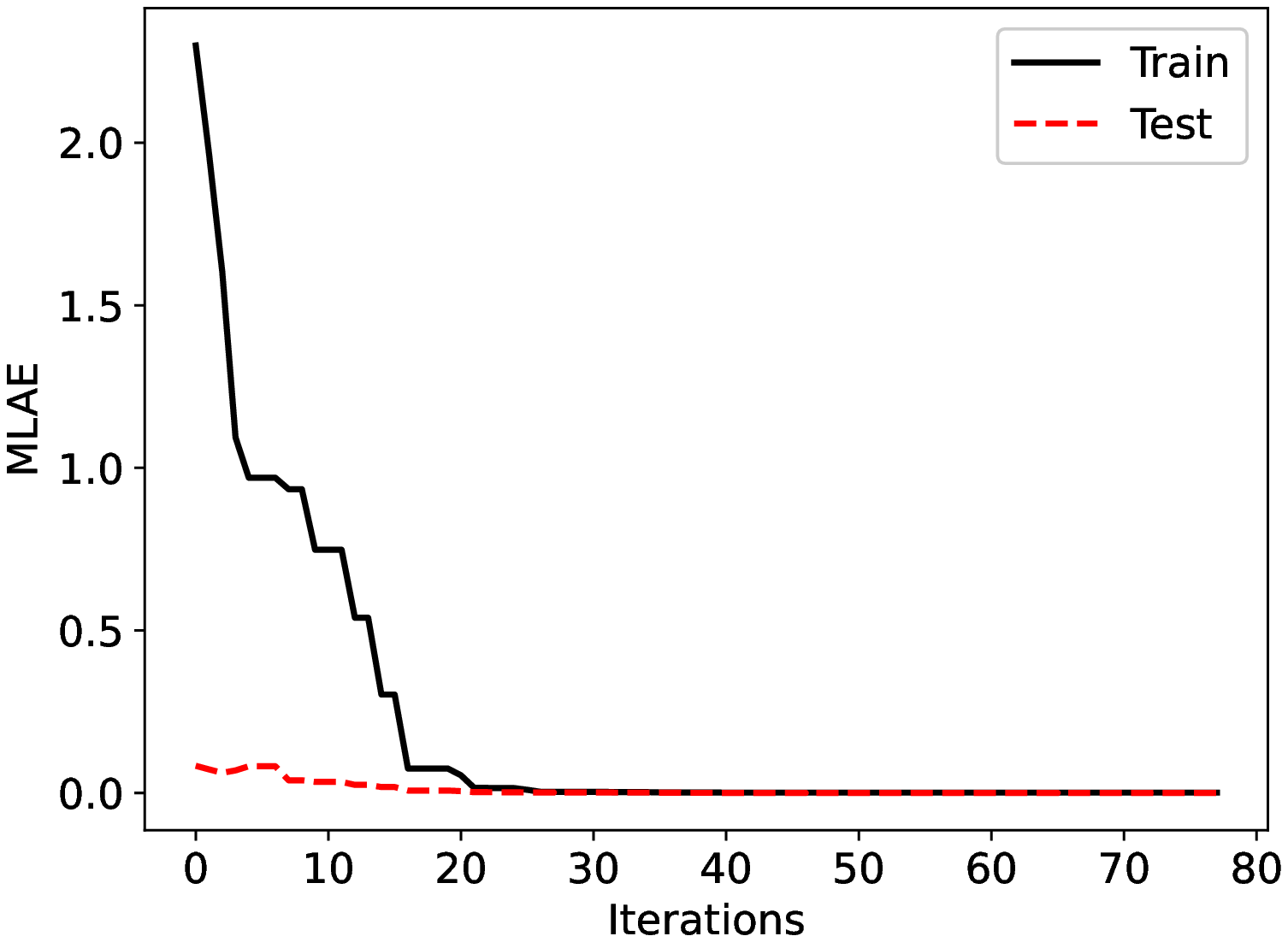}}
	\subfloat[Solution path.]{\includegraphics[width=0.4\textwidth]{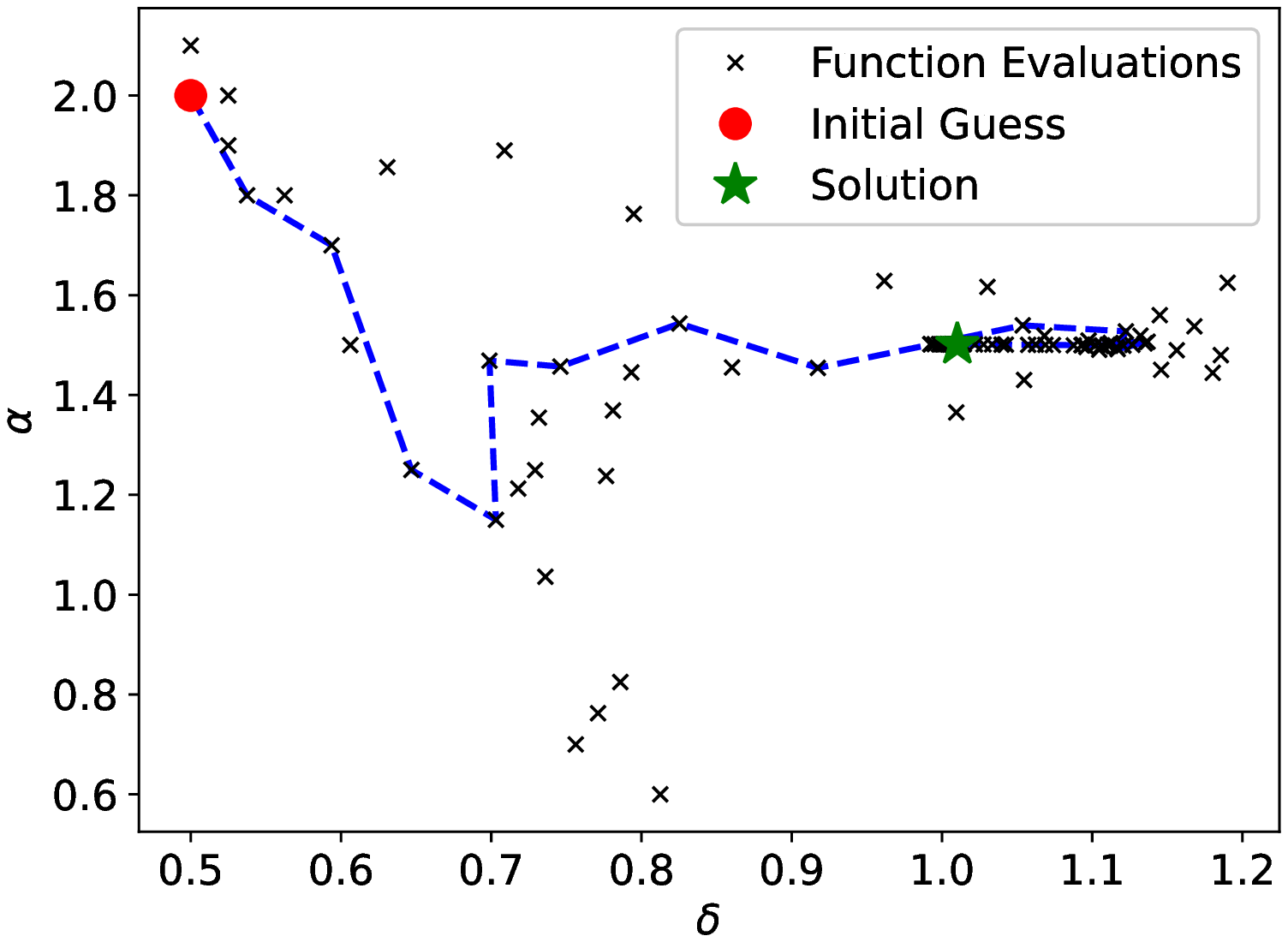}}
	\caption{Iteration errors (training and testing), and the solution path of the ML algorithm when solving the inverse problem of a manufactured solution with a known kernel.}
	\label{nlml_fig:manuf}
\end{figure}

We plot the training and test errors over the number of iterations in Fig~\ref{nlml_fig:manuf}. We also illustrate the solution path from the initial guess  to the final parameter estimates of $\alpha$ and $\delta$, explicitly showing the function evaluations driving the iterations of Nelder-Mead optimization. We further explore the robustness of the algorithm with the DDD-based dataset.

\subsection{DDD-Driven Results}

We now employ the ML framework on the dataset generated by DDD simulations, represented by the PDFs of shifted dislocation positions obtained at the last 10000 time-steps, as highlighted in Fig.~\ref{nlml_fig:sel_data}. We expand the robustness assessment of the framework, and we test other critical aspects such as sensitivity to the initial guess and train/test ratios. 

The AKDE algorithm removes the noise from lack of data-points, especially at the tails, and produces a smooth curve throughout the domain. The intrinsic noise related to the variable number of particles is embedded in the PDF estimation, leading to smooth curves throughout the time-range of our data. For these reasons, we expect the algorithm to perform well with the DDD data.

We start by investigating the solution with different train-test splits. We compare the results of a $80/20$ split as in the manufactured solution with a $60/40$ split. We adopt an initial guess of $\alpha_0 = 2$, $\delta_0 = L/2$, which resides at the center of the same parametric range used before, $\alpha = [0,4]$, and $\delta = [\Delta x,L]$. We run Algorithm~\ref{nlml_algo:NM} for Cases 1, 2, and 3, and collect the optimal values of $\alpha$, $\delta$, $D$, the computational cost in terms of Nelder-Mead iterations, the cost function evaluations, and the training and validation cost. We present those results in Table~\ref{nlml_tab:split}.

We highlight the results of $\alpha$ in Table~\ref{nlml_tab:split} in comparison to the exponent of power-lay scaling in velocity distributions from Fig.~(\ref{nlml_fig:dist}). We note that the faster velocity decay in Case 1 with $\beta = 3$ is consistent with the kernel exponent $\alpha = 2.99$. Similarly, for Cases 2 and 3, the velocity distribution decay was found to be around $\beta = 2.4$, while the kernel exponent from the ML was found to be $\alpha = 2.40$ for Case 2 and $\alpha = 2.54$ for Case 3 under the $80/20$ split. We will further comment this connection on the Discussion section.

The results obtained with the two train-test splits are equivalent in terms of the kernel parameters and the overall cost. Indeed, there is no evident difference in choosing one ratio over the other. The main difference comes in the overall training and testing cost. We observe that in all cases the training cost is larger than the test cost, similarly to the results obtained with the manufactured solution. This is due to the time-dependent dynamics of the PDF evolution and the higher availability of training points, as in the manufactured case. However, here we have another contributing factor. As discussed in Section~\ref{sec:data}, earlier data-points will be heavily influenced by the initial load application, while late points will be closer to a steady-state. The test set in the $60/40$ split includes more earlier points, and therefore sees their influence reflected in higher training costs, besides allowing for less training points to make the model more general. For the remaining results, we choose the $80/20$ split as our representative case.

\begin{table}[]
	\centering
	\caption{Machine Learning results for two train-test split solutions.}
	\label{nlml_tab:split}
	\begin{tabular}{lllllll} \toprule
		& \multicolumn{2}{c}{Case 1} & \multicolumn{2}{c}{Case 2} & \multicolumn{2}{c}{Case 3} \\ 
		Train/Test Split  & 80/20        & 60/40       & 80/20        & 60/40       & 80/20        & 60/40       \\ \midrule
		$\alpha$          & 2.99         & 3.00           & 2.40          & 2.37        & 2.54         & 2.51        \\
		$\delta$          & 20.62        & 18.79       & 33.7         & 33.07       & 34.03        & 33.72       \\
		$D$          & 3.64e-4       & 3.55e-4       & 7.40e-4         & 7.66e-4       & 1.78e-3        & 1.83e-3       \\
		\# Iterations     & 79           & 75          & 46           & 79          & 60           & 85          \\
		\# Evaluations & 154          & 149         & 89           & 157         & 108          & 159         \\
		Training Cost    & 6.30e-02     & 5.73e-02    & 6.39e-02     & 5.55e-02    & 6.75e-02     & 5.57e-02    \\
		Testing Cost     & 4.42e-02     & 6.33e-02    & 4.95e-02     & 8.70e-02    & 4.28e-02     & 6.21e-02   \\ \bottomrule
	\end{tabular}
\end{table}

\begin{figure}[]
	\centering
	\subfloat[Case 1.]{\includegraphics[width=0.33\textwidth]{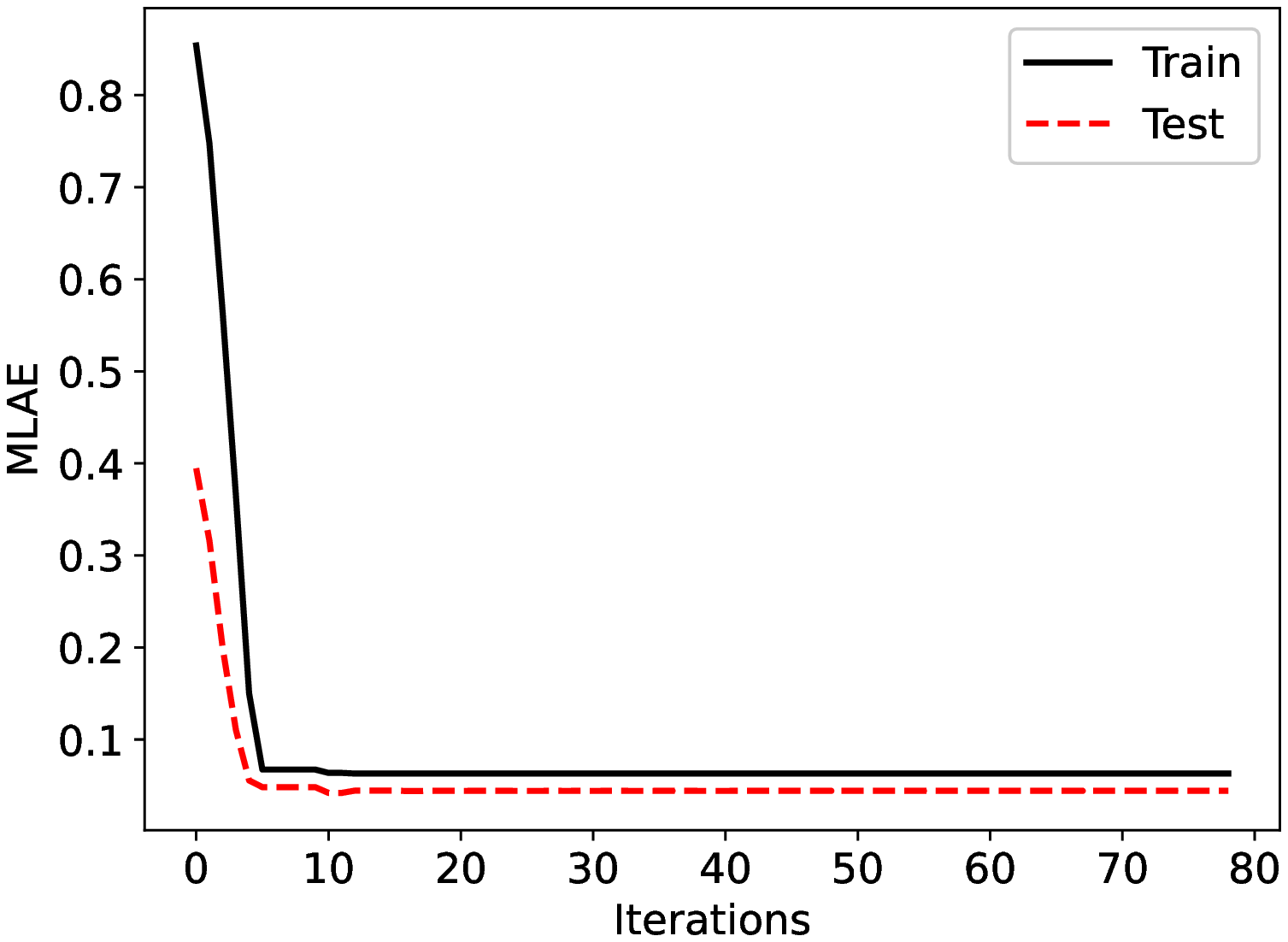}}
	\subfloat[Case 2.]{\includegraphics[width=0.33\textwidth]{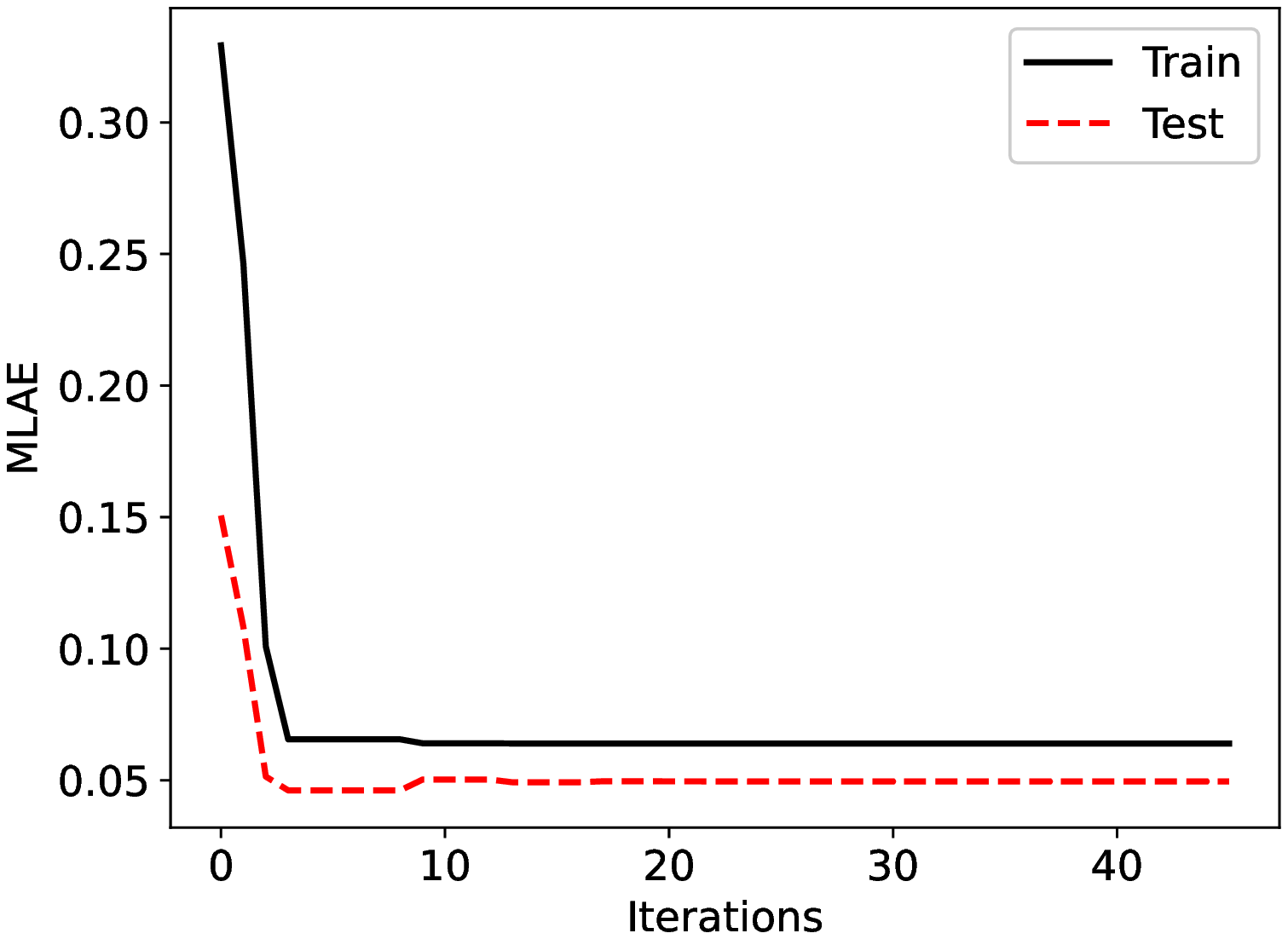}}
	\subfloat[Case 3.]{\includegraphics[width=0.33\textwidth]{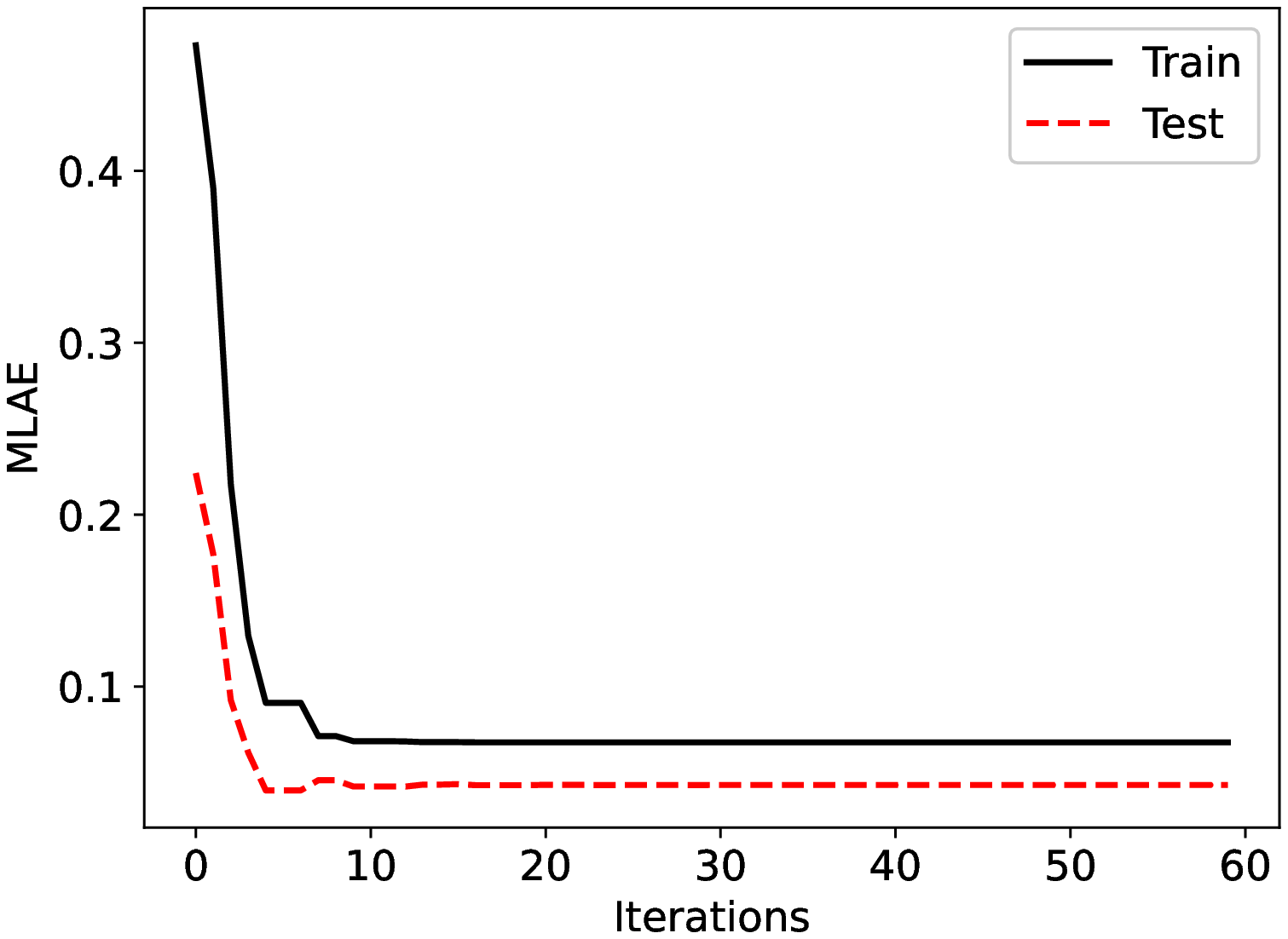}}
	\caption{Evolution of training and testing MLAE values computed over the NM iterations.}
	\label{nlml_fig:cost}
\end{figure}

We plot the evolution of training and test MLAE values over the Nelder-Mead iterations in Fig~\ref{nlml_fig:cost}. We see that the algorithm quickly gets near the solution as the errors drop sharply near the initial iterations. Then, the errors remain nearly constant as the minimizer further explores the research space in the proximity of the minimum.  

We further explore the capabilities of the proposed ML algorithm and test the performance of the parameter learning under different initial conditions beyond the central point, and choose four extra points near the corners of our parametric search space:

\begin{enumerate}
	\item Guess 1 (original guess at the center): $\alpha_0 = 2$, $\delta_0 = L/2$.
	\item Guess 2: $\alpha-0 = 1$, $\delta_0 = L/4$.
	\item Guess 3: $\alpha_0 = 3$, $\delta_0 = 3L/4$.
	\item Guess 4: $\alpha_0 = 3$, $\delta_0 = L/4$.
	\item Guess 5: $\alpha_0 = 1$, $\delta_0 = 3L/4$.
\end{enumerate}

\begin{figure}[]
	\centering
	\subfloat[Case 1.]{\includegraphics[width=0.33\textwidth]{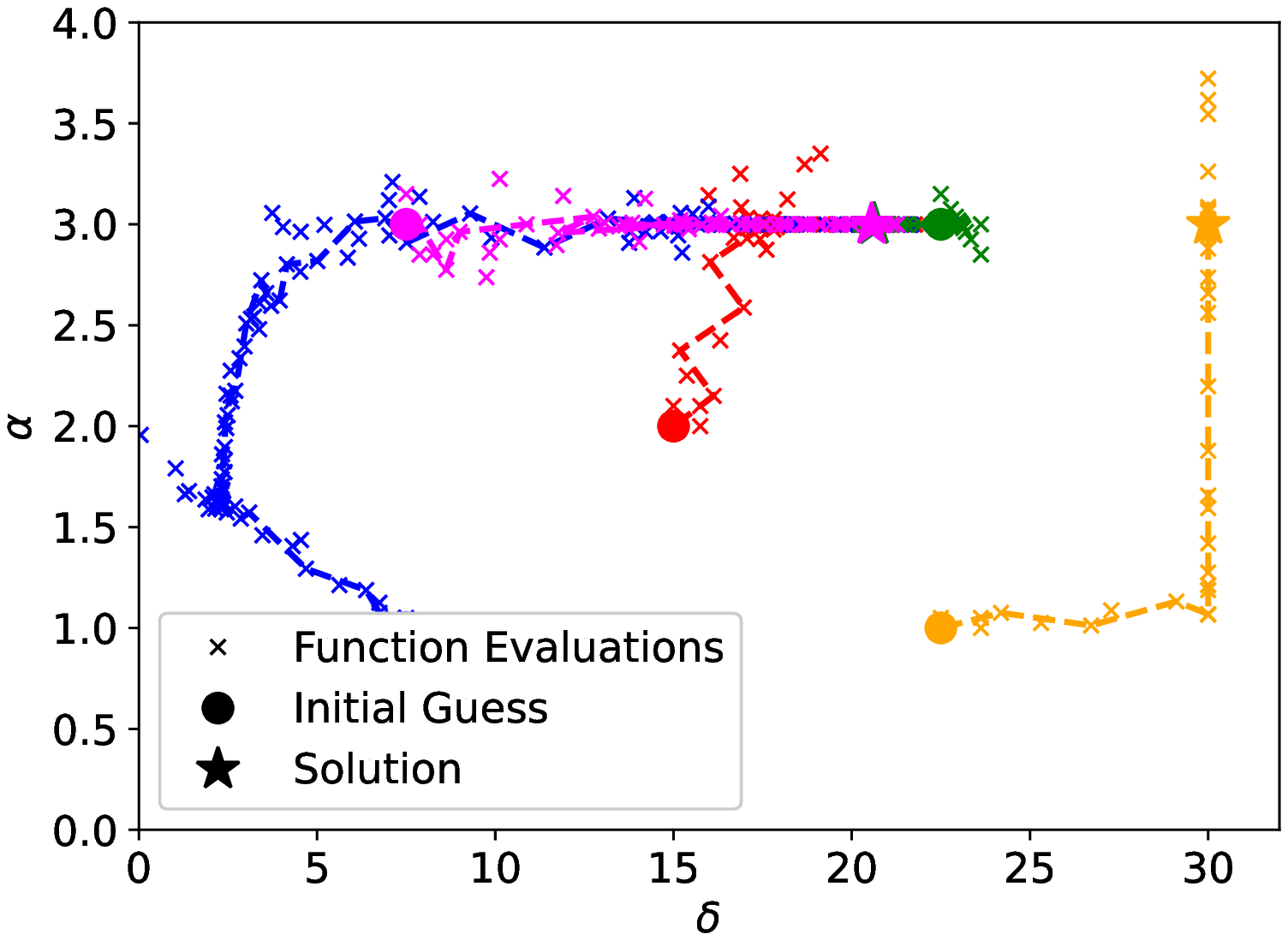}}
	\subfloat[Case 2.]{\includegraphics[width=0.33\textwidth]{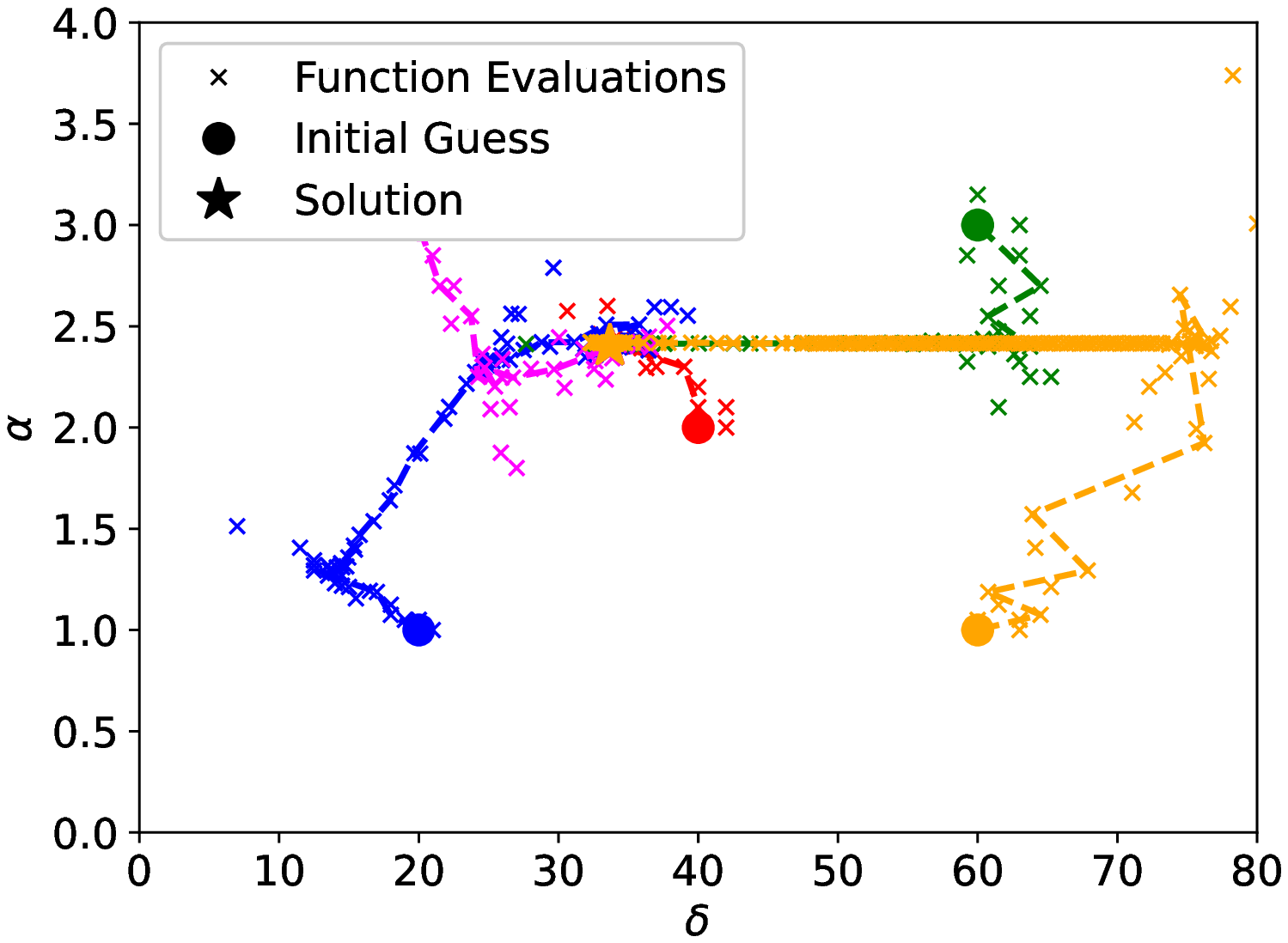}}
	\subfloat[Case 3.]{\includegraphics[width=0.33\textwidth]{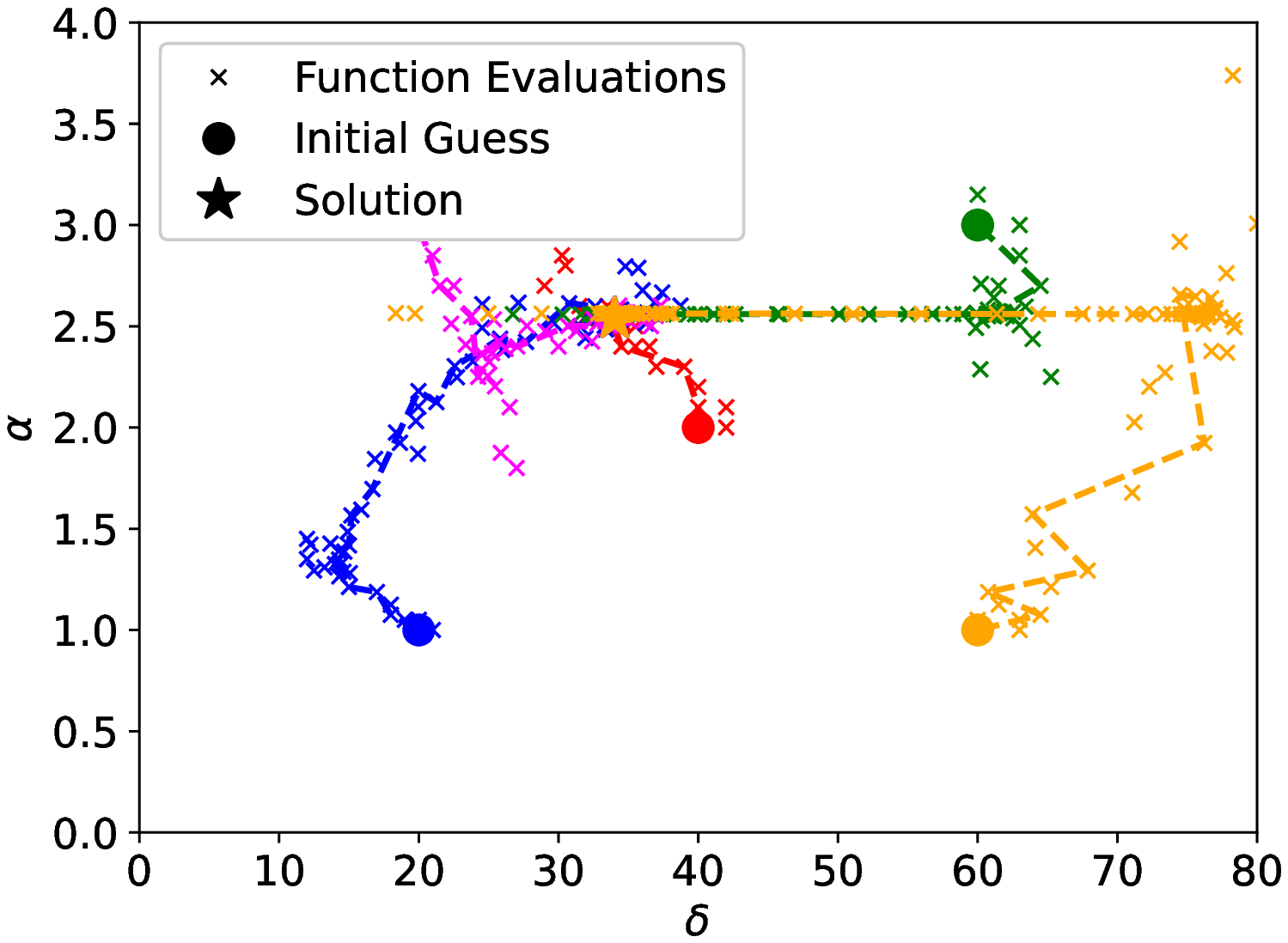}}
	\caption{Solution path from different combinations of initial guess.}
	\label{nlml_fig:guess}
\end{figure}

We present the final results in Table~\ref{nlml_tab:guess}. In general, an initial guess close to the center of the parametric space leads to less iterations for Cases 2 and 3.  The only different result we obtain is for Case 1, Guess 5, where the horizon $\delta$ is computed as the upper-bound $L$, yet with $\alpha$, $D$, and MLAE values sufficiently close to the results of other initial guess combinations. Based on this observation, the horizon $\delta$ seems to have a lower bound, above which the results are less sensitive to increasing horizon. We show the different paths the algorithm takes under the proposed initial guess combinations in Fig.~\ref{nlml_fig:guess}, where we can see the function evaluations made by the algorithm and the final solutions, illustrating their proximity. We can also distinguish the upper-bound solution of Case 1, Guess 5 in the same figure.

\begin{table}[]
	\centering
	\caption{Machine Learning results for different initial guess combinations of $\alpha$ and $\delta$.}
	\label{nlml_tab:guess}
	\begin{tabular}{cllllll} \toprule
		\multicolumn{1}{l}{Case} &                   & Guess 1 & Guess 2 & Guess 3 & Guess 4 & Guess 5 \\ \midrule
		\multirow{4}{*}{1}       & $\alpha$          & 2.99    & 2.99    & 2.99    & 2.99    & 2.99    \\
		& $\delta$          & 20.62   & 20.63   & 20.63   & 20.56   & 30.00   \\
		
		& $D$          & 3.63e-4   & 3.63e-4   & 3.63e-4   & 3.63e-4   & 3.63e-4  \\
		& \# Iterations     & 79      & 122     & 57      & 76      & 30      \\
		& \# Evaluations & 154     & 219     & 114     & 152     & 59      \\ \midrule
		\multirow{4}{*}{2}       & $\alpha$          & 2.40    & 2.40    & 2.40    & 2.40    & 2.40    \\
		& $\delta$          & 33.70   & 33.66   & 33.66   & 33.68   & 33.68   \\
		& $D$          & 7.40e-4   & 7.40e-4   & 7.40e-4   & 7.40e-4   & 7.40e-4  \\
		& \# Iterations     & 46      & 73      & 86      & 67      & 129     \\
		& \# Evaluations & 89      & 142     & 163     & 132     & 253     \\ \midrule
		\multirow{4}{*}{3}       & $\alpha$          & 2.54    & 2.54    & 2.54    & 2.54    &    2.54     \\
		& $\delta$          & 34.03   & 34.02   & 34.03   & 34.03   &   34.01      \\
		& $D$          & 1.78e-3   & 1.78e-3   & 1.78e-3   & 1.78e-3   & 1.78e-3   \\
		& \# Iterations     & 60      & 87      & 120     & 68      &      211   \\
		& \# Evaluations & 108     & 169     & 233     & 126     &   408     \\ \bottomrule
	\end{tabular}
\end{table}

\begin{figure}[H]
	\centering
	\includegraphics[width=0.4\textwidth]{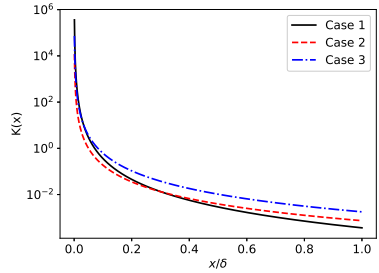}
	\caption{Final kernel shapes from optimized parameters obtained trough the ML algorithm, scaled by $\delta$.}
	\label{nlml_fig:kernels}
\end{figure}

We choose the results from Guess 1, at the center of the parametric space, to be the representative parameters for kernel reconstruction and visualization. From the optimal values of the power-law decay exponent $\alpha$, horizon $\delta$, and coefficient $D$, we compute the nonlocal kernel following the definition from Eq.~(\ref{nlml_eq:kernel}). We plot the kernel shapes for Case 1, 2, and 3 in Fig~\ref{nlml_fig:kernels}. 

Finally, to illustrate the nonlocal model's potential to simulate the evolution of position PDFs from DDD simulations, we run the model using Eq.~(\ref{nlml_eq:full_discrete}) with the optimal parameter values from Guess 1 combination, starting from the initial time-step of training, until the last time-step of testing, covering the whole range of available data. We measure the accuracy of the model using the $l_2$ relative error at the last time-step, defined as

\begin{equation}
\epsilon = \frac{\Vert \tilde{p} - p \Vert_{2}}{\Vert p \Vert_2},
\end{equation}

\noindent where $\tilde{p}$ represents the model solution at the specified time-step, and $p$ is the true PDF obtained from DDD at the same time measure. 

\begin{figure}[]
	\centering
	\subfloat[Case 1.]{\includegraphics[width=0.33\textwidth]{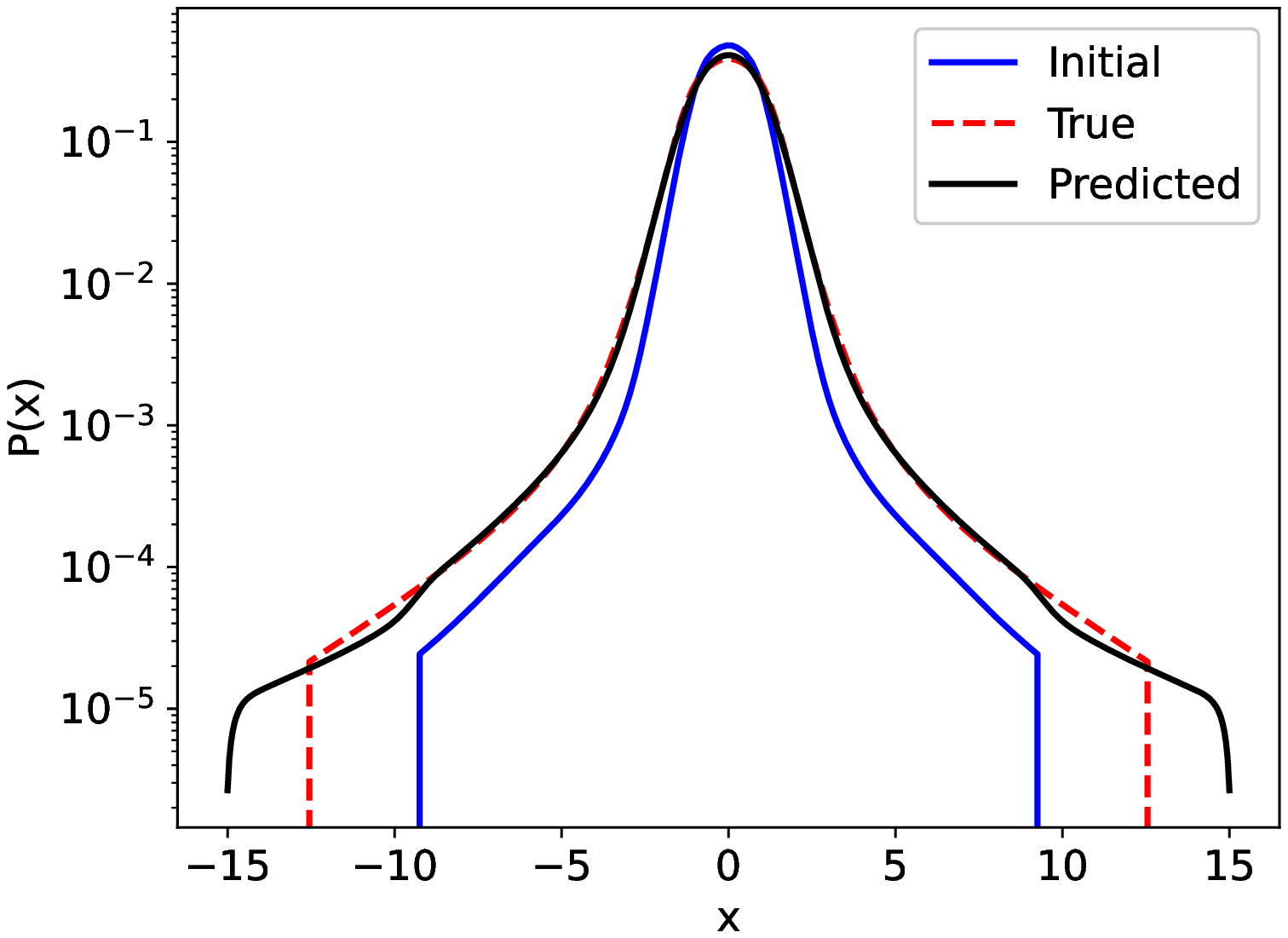}}
	\subfloat[Case 2.]{\includegraphics[width=0.33\textwidth]{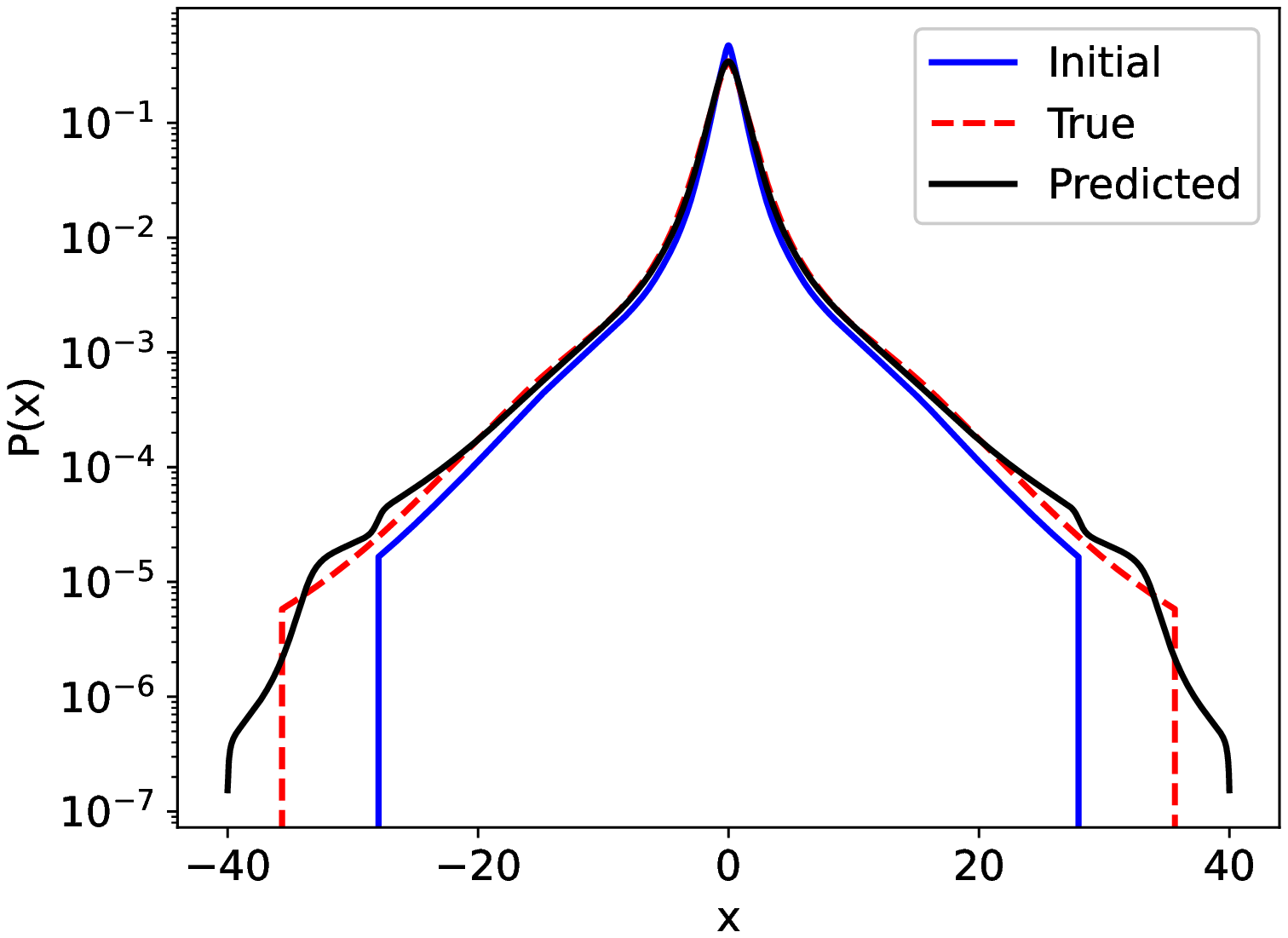}}
	\subfloat[Case 3.]{\includegraphics[width=0.33\textwidth]{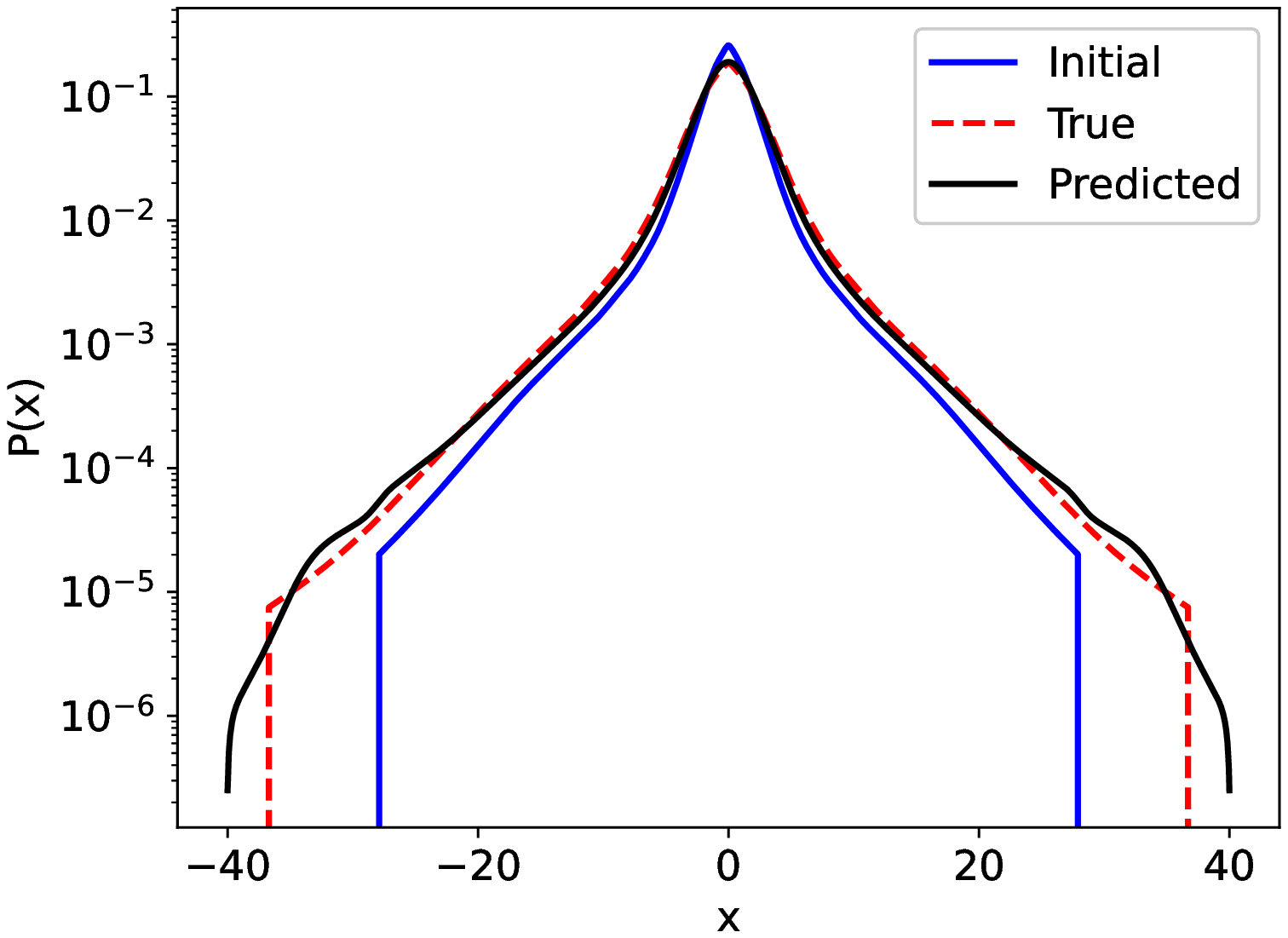}}
	\caption{Simulation of the nonlocal model for the whole time-interval of available data, highlighting the initial PDF, and the final distributions of the true data and the nonlocal prediction.}
	\label{nlml_fig:final_pdf}
\end{figure}

We compute the relative $l_2$ error and obtain $\epsilon_1 = 4.75\text{e-}2$, $\epsilon_2 = 4.22\text{e-}2$, and $\epsilon_3 = 6.30\text{e-}2$ for Case 1, Case 2, and Case 3, respectively. Considering that this simulation takes over the 10000 time-steps of available data, the maximum relative error of $6.3\%$ in the $l_2$ sense for Case 3 shows that the model can successfully reproduce the overall dynamics of the fluid-limit motion of dislocation particles in one dimension. We further illustrate the final shape of the PDF from the model, and compare it with the true shape at the final time-step in Fig~\ref{nlml_fig:final_pdf}. 

We verify that the ML algorithm successfully captured the parameters that best describe the evolution of dislocation position PDFs according to the proposed nonlocal diffusion model. 

\subsection{Discussion}

Given the broad scope of this work, we divide the main discussion among three dominant facets. We discuss the overall capabilities of the proposed bi-level ML algorithm, followed by a discussion on the nonlocal model itself, and how the nonlocal kernel is connected to the particle dynamics observed in DDD.

We start by examining the ML aspects of the data-driven approach. It is evident that the proposed framework for solving the inverse problem of finding the parameters of a nonlocal Laplace operator is successful in the present scenario. Starting with the manufactured solution, we see the training error quickly approaching the plateau in Fig.~(\ref{nlml_fig:manuf}. The decoupling of the nonlocal diffusion coefficient $D$ from other kernel parameters $\alpha$ and $\delta$ (which is possible due to the linearity of the operator) facilitates the dimensionality reduction and the implementation of the bi-level algorithm.

The robustness is guaranteed by the data-driven learning of a DDD-based kernel. The bi-dimensional optimization algorithm converged in as few as $O(10^2)$ function evaluations and $O(10)$ Nelder-Mead iterations in most cases. We also observed the convergence to the same parameters in all but one different combinations of initial guess. The use of high-performance computing and parallelization further enhanced the performance of the learning framework, making it scalable. Additionally, the proposed framework easily generalizes to more complex kernel shapes other than a pure power-law decaying shape, by simply adding the multiplicative factor $P$ and learning its parameters. In fact, when defining the kernel as, e.g., a combination of basis functions, the algorithm seamlessly accommodates this approximation through inserting additional columns on the RHS of Eq.~(\ref{nlml_eq:matrix}), and solving for more coefficients at Level 1.

The efficiency of the algorithm is reflected in the consistency of results obtained in Cases 1, 2, and 3. It is clear that the nonlocal model is the appropriate choice for this particular problem, evidenced by the large value of horizon $\delta$, to the order of $400$-$600$ times the grid size. The long-range interactions from DDD are therefore represented as a nonlocal kernel with large horizon. Moreover, multiplication mechanisms correspond to larger values of $\delta$, since the avalanches and the associated collective dynamics, represented by intermittency in velocity signals, lead to heavy-tail velocity distributions, which translates into heavy tails in the corresponding PDFs. This is in contrast with Case 1 without multiplication, where we have the PDFs closer to normal distributions than the ones from Cases 2 and 3. Therefore, the anomalous behavior in the discrete case leads to nonlocality also in the continuum case.

The other immediate observation related to the nonlocal model is the meaning of $\alpha$, which clearly distinguishes the dynamics of dislocations with and without multiplication. For Case 1 without multiplication, we obtain a value of $\alpha$ closer of $2.99$, while the multiplication mechanisms of Cases 2 and 3 are translated into $\alpha = 2.4$ and $\alpha = 2.54$, respectively. As anticipated in Section \ref{nlml_sec:nonlocal}, the operator $\mathcal L$ obtained with the power-law nonlocal kernel with finite horizon is equivalent to the truncated fractional Laplacian of fractional order $s$ via the relationship $\alpha=1+2s$. Under such view, Case 1 would correspond to a fractional Laplacian with $s = 0.99$, while Cases 2 and 3 would take $s = 0.7$ and $s = 0.77$, respectively. In this perspective, it is straightforward to see the evolution of dislocation PDFs as being super-diffusive, with Case 1 being the closest to a classical diffusion process, yet with a pronounced nonlocality due to rearrangements in the dislocation structure due to annihilations. However, the multiplication mechanism is the main factor that turns a rather diffusive process into super-diffusive. Moreover, the super-diffusion is intensified under a lower load, Case 2, where the external stress state allows faster relaxation and stronger subsequent avalanches compared to Case 3, where the overall higher stress state makes all dislocations move faster, yielding less relaxation time to a critical, metastable configuration.

The most striking observation related to the kernel discovery is the correspondence between the kernel fractional parameter $\alpha$ and the scalings of velocity distribution tail from Fig.~\ref{nlml_fig:dist}. The empirical scaling observed here and among other works in the literature matches the values of $\alpha$ found by the data-driven kernel learning algorithm. This is not surprising, as we can take the velocity distribution to be jump size distributions that define a particular L\'evy measure, essential when transforming a stochastic process into differential equations through the Semigroup theory \cite{meerschaert2011stochastic}. The formal and complete definition of the stochastic process that governs the dislocation trajectories is out of the scope of this paper, yet we see jump size distributions following Fig.~\ref{nlml_fig:dist} lead to operators defined by power-law kernels taking the form of Eq.~(\ref{nlml_eq:kernel}).

From a wider perspective, the procedure and reasoning developed in this paper do not need to be restricted to dislocation dynamics. The problem of learning kernels and dynamics from high-fidelity simulations and real-life data is a relevant research topic of increasing popularity. While the methodology presented in this work highlights the inference of a nonlocal operator from physical mechanisms, more specifically dislocation dynamics, the same procedure may be applied to tie the use of other non-standard nonlocal operators to the anomalous physical processes that they describe. 

\section{Conclusions}
\label{sec:conclusions}

We proposed a data-driven nonlocal model for the simulation of dislocation position probability densities. We generated dislocation shifted position data in the form of particle trajectories from high-fidelity two-dimensional DDD simulations under creep with different load levels, with and without multiplication mechanisms. From the Lagrangian particle trajectories we estimated the evolution of PDFs through and Adaptive Kernel Density Estimation method. Last, we developed a bi-level ML algorithm to obtain the kernel parameterization for the proposed nonlocal operator that describes that PDF's evolution. The developed approach integrates the high-fidelity dynamics of dislocations at the meso-scale with a continuum probabilistic frame in a fluid-limit sense.

We make the following observations from the integrated framework:

\begin{itemize}
	\item We recovered the dislocation velocity statistics available in the literature from our two-dimensional DDD simulations. We identified the same exponent of around $2.4$ for the power-law decay velocity distribution tail when dislocation multiplication is present, and a sharper, close to $3$ exponent without the effect of multiplication mechanisms. The statistics from DDD show similarities among the studied cases.
	\item The PDF estimation from dislocation trajectories makes evident that the presence of multiplication sources greatly impacts the probability distributions, increasing the heaviness of the tails and implying greater nonlocality. 
	\item Our bi-level algorithm based on a Least-Squares approach for the computation of the optimal nonlocal diffusion coefficient for every pair of $\alpha$ and $\delta$ performed well with a manufactured solution, and proved to be robust for data-driven PDFs, considering different train-test splits and initial guess combinations.
	\item The large horizon parameter $\delta$ found among the cases confirms the nonlocal nature of dislocation dynamics, even from a probabilistic perspective. Furthermore, the nonlocal kernel power-law exponent obtained matches the tail decay from dislocation velocity distributions computed in DDD simulations. This establishes a well-defined path between the anomalous behavior observed in particle meso-scale dynamics and the upscaling of anomalous effects to a continuum, macro-scale frame of reference.
\end{itemize}

Although we used single-glide mechanisms, we note that shifted particle positions may be obtained in more general, multi-slip systems. Since the goal of this framework is to obtain the probability densities in the fluid-limit, the same procedure could be applied to each slip-system in a complex crystal. We further point to the fact that the bulk dynamics adopted here can also be extended to dislocation motion near free surfaces, crack tips, or grain boundaries. In such cases, one could expect the PDFs to show non-zero skewness, which could be naturally accommodated by a different choice of (nonsymmetric) kernel in the nonlocal operator, suitable for skewed, possibly one-sided distributions.

The nonlocal model of dislocation motion at the meso-scale proposed in this work opens up the opportunity of fast computations of quantities of interest compared to the high-fidelity simulations. The implications of nonlocal dislocation models are readily applicable to the study of visco-elasticity and visco-plasticity, where fractional-order models have been successfully applied to model the power-law relaxation including damage effects \cite{suzuki2021thermodynamically}. One of the main connections to be established is the ultimate effect of different regimes of dislocation dynamics on the evolution of macro-scale free-energy potentials during failure, in phase-field models \cite{barros2021integrated} for instance. Around crack tips and other dislocation generation objects, such as holes, pores, or other micro-cracks, we expect the macro-scale behavior to also be anomalous. Substantially, the methods proposed here can be essential tools to connect other physical processes from a wider range of applications to the generation of corresponding nonlocal operators.

Finally, the proposed bi-level optimization approach is an effective way of reducing the computational burden of optimizing in a high-dimensional parameter space and proves to be robust with both manufactured and simulated datasets.

\bibliographystyle{siamplain}
\bibliography{reference}

\end{document}